\documentclass[aps,prx,amsmath,floats,floatfix,twocolumn,superscriptaddress,nofootinbib,showpacs]{revtex4-1}

\usepackage{amssymb,amsmath,verbatim,mathtools,needspace,enumitem,etoolbox,graphicx,microtype,afterpage,bm,revsymb}
\usepackage{mathrsfs}

\usepackage[dvipsnames, usenames]{xcolor}
\definecolor{linkcolor}{rgb}{0.1,0.2,0.6}
\usepackage[unicode, colorlinks=true, linkcolor=linkcolor, citecolor=linkcolor, filecolor=linkcolor,urlcolor=linkcolor, pdfusetitle]{hyperref}

\usepackage[T1]{fontenc}
\usepackage[utf8]{inputenc}
\usepackage{tabularx, booktabs}
\usepackage[italicdiff]{physics}
\usepackage{float}
\usepackage[caption=false]{subfig}
\usepackage[normalem]{ulem}

\graphicspath{{./Figures/}}

\begin{document}

\title{Analytic and accurate approximate metrics for black holes with arbitrary rotation \\ in beyond-Einstein gravity using spectral methods}

\author{Kelvin~Ka-Ho~Lam} 
\email{khlam4@illinois.edu}
\affiliation{Illinois Center for Advanced Studies of the Universe \& Department of Physics, University of Illinois Urbana-Champaign, Urbana, Illinois 61801, USA}

\author{Adrian Ka-Wai Chung}
\email{akwchung@illinois.edu}
\email{kwc43@cam.ac.uk}
\affiliation{Illinois Center for Advanced Studies of the Universe \& Department of Physics, University of Illinois Urbana-Champaign, Urbana, Illinois 61801, USA}
\affiliation{DAMTP, Centre for Mathematical Sciences, University of Cambridge, Wilberforce Road, Cambridge CB3 0WA, United Kingdom}

\author{Nicol\'as Yunes}
\affiliation{Illinois Center for Advanced Studies of the Universe \& Department of Physics, University of Illinois Urbana-Champaign, Urbana, Illinois 61801, USA}

\date{\today}

\begin{abstract}

    A key obstacle for theory-specific tests of general relativity is the lack of accurate black-hole solutions in beyond-Einstein theories, especially for moderate to high spins. 
    We address this by developing a general framework--based on spectral and pseudospectral methods--to obtain analytic, closed-form spacetimes representing stationary, axisymmetric black holes in effective-field-theory extensions of general relativity to leading order in the coupling constants. 
    The approach models the spacetime (and extra fields) as a stationary, axisymmetric deformation of the Kerr metric in Boyer-Lindquist-like coordinates, expands metric deformations as a spectral series in radius and polar angle, and converts the resulting beyond-Einstein field equations into algebraic equations for the spectral coefficients. 
    For any given spin, these equations are solved via standard linear-algebra methods; the coefficients are then fitted as functions of spin with non-linear functions, yielding fully analytic metrics for rotating black holes in beyond-Einstein theories. 
    We apply this to quadratic gravity theories--dynamical Chern-Simons, scalar-Gauss-Bonnet, and axi-dilaton gravity--obtaining solutions valid for any spin, including near-extremal cases with errors below machine precision for $a \leq 0.9$ and $\lesssim 10^{-8}$ for $a \leq 0.999$. 
    We show that existing slowly-rotating solutions break down at $a \sim (0.2, 0.6)$, depending on approximation order and chosen accuracy. 
    We then use our metrics to compute observables, such as the surface gravity, horizon angular velocity, and the locations of the innermost circular orbit and the photon ring. 
    The framework is general and applicable to other effective-field-theory extensions for black holes of any spin.
    
\end{abstract}

\maketitle

\section{Introduction}\label{sec:introduction}

General relativity (GR) is one of the most remarkable theories in physics to date. 
Over the past century, GR has passed numerous tests in the Solar System and binary pulsars \cite{Will2014, Yunes:2013dva}. 
Einstein's theory, however, is famously known to be incompatible with quantum physics~\cite{tHooft:1974toh, Deser:1974cz, Deser:1974xq}, and by itself is not able to explain cosmic inflation \cite{Guth:1980zm, Starobinsky:1980te}, black hole (BH) singularities \cite{Hawking:1970zqf}, or galaxy rotation curves \cite{Sofue:2000jx, Rubin:1970zza} without invoking the dark matter \cite{Bertone:2018krk}. 
These open questions have recently motivated the study of beyond-Einstein theories that, in particular, could be tested against the recent flood of gravitational wave data \cite{LIGOScientific:2025slb, LIGOScientific:2021djp, LIGOScientific:2020ibl}. 
This data contains, in principle, rich information about the ``extreme gravity regime,'' where the spacetime curvature is simultaneously large and dynamical, which is probed when BHs collide and neutron stars merge. 

The Lagrangian density of beyond-Einstein gravity differs from that of the Einstein-Hilbert action.
A natural extension of the latter is to include higher powers of the spacetime curvature, perhaps multiplied by functions of a scalar or vector field, to the Lagrangian density \cite{Will2014}.
Truncating this expansion at a given order in curvature defines a wide class of beyond-Einstein effective field theories (EFTs) that are valid below a certain curvature or energy scale. 
For example, the quadratic-order family of beyond-Einstein theories that contains scalar fields (henceforth, quadratic gravity theories for short) includes \cite{Yunes:2011we, Cano_Ruiperez_2019} scalar-Gauss-Bonnet (sGB) gravity \cite{Yagi:2015oca, Nair:2019iur}, dynamical Chern-Simons (dCS) gravity \cite{dCS_01, dCS_02} and axi-dilaton (AD) gravity \cite{Kanti:1995cp, Cano:2021rey}.
Theories of this type have seen a resurgence lately~\cite{Doneva:2021dcc, Alexander:2024vav, Alexander:2021ssr, Meng:2023wgi, Perkins:2021mhb, Cano:2021rey} because they are well-motivated (i.e.~they emerge as the low-energy limit of different grand unification frameworks, such as string theory \cite{Maeda:2009uy, Moura:2006pz, Cano_Ruiperez_2019, Cano:2021rey} or loop quantum gravity \cite{dCS_01, Taveras:2008yf}) and they can lead to potentially observable signatures in various current experiments~\cite{Berti:2015itd, Berti:2025hly, Shankaranarayanan:2022wbx}. 

Rotating BHs in beyond-Einstein gravity may differ from those in GR, as is certainly the case in quadratic gravity. 
For example, in dCS gravity, while the Schwarzschild metric remains a valid solution to the dCS field equations for spherically symmetric and static, vacuum spacetimes, the Kerr metric is not a solution for stationary and axisymmetric, vacuum spacetimes \cite{Grumiller:2007rv}. 
Likewise, BH solutions in sGB or AD gravity differ from the Schwarzschild and Kerr solutions of GR \cite{Cano_Ruiperez_2019}. 
In order to compute physical observables associated with beyond-Einstein BHs (be it observables related to gravitational waves or accretion disks, for example) that can be compared to data, one must first solve the modified field equations to obtain the exterior BH spacetime. 
This has proven to be extraordinarily difficult in quadratic gravity theories when searching for analytic (i.e.~closed-form) or numerical and accurate expressions of beyond-Einstein BHs with any rotation. 

The main difficulty is that beyond-Einstein field equations are typically a coupled set of partial differential equations that is even more nonlinear than the Einstein equations. Additionally, if there are dynamical scalar or vector fields present in the action, then these will couple non-trivially to the metric functions. For example, in quadratic gravity, a scalar (or a pseudo-scalar) field $\varphi$ couples to a parity-even (or a parity-odd) quadratic curvature invariant $\mathscr{Q}$ in the action, multiplied by a coupling constant $\alpha$. Since the field is dynamical, its canonical kinetic term yields a wave-like equation of motion that is sourced by the quadratic curvature invariant. Meanwhile, the variation of the action with respect to the metric yields field equations that look like Einstein's, but that contain two sources: the stress-energy tensor of the field and a highly-nonlinear term generated from the variation of $\mathscr{Q}$ with respect to the metric. As one can imagine, solving this coupled set of partial differential equations is exceedingly difficult. 

A way forward, nonetheless, has emerged by combining two facts. The first fact is that the beyond-Einstein gravity theories described above are well-motivated only as \textit{EFTs}, and thus, they possess a limited regime of validity. In practice, this means there is a cut-off curvature (or energy) scale above which the effective theory ceases to be valid, because additional higher-curvature terms would have to be included above this scale. For quadratic curvature theories and BH spacetimes, this means one must restrict attention to systems in which the coupling constant $\alpha$ and the BH mass $M$ satisfy $\alpha/M^2 \ll 1$, which has come to be known as the ``small-coupling approximation.'' The second fact is that existing tests of GR indicate that any beyond-Einstein gravity effect, if present, must be \textit{small} relative to GR \cite{Yunes:2013dva}. If this were not the case, then it would be highly likely that the plethora of GR tests that have already been carried out (e.g.~with binary pulsars and with gravitational-wave observations) would have hinted toward a GR deviation. Since no GR deviations in extreme gravity observations have been found to date, any deviations that may be hiding in the noise must be small relative to the predictions of Einstein's theory.   

These facts allow us to attack the problem of finding beyond-Einstein, rotating BH solutions through a perturbative, small-coupling scheme, i.e., expanding the fields as a series in the coupling constant $\alpha$. In this scheme, one typically solves the scalar or pseudo-scalar equation of motion first, assuming a GR background. One then linearizes the beyond-Einstein field equations in the metric deformation and uses the previous scalar or pseudo-scalar field solution to solve for the metric deformation. This approximation is equivalent to the EFT framework in which the field equations, at each order in $\alpha$, must be satisfied independently \cite{Shankaranarayanan:2022wbx}. 
In this limit, quadratic gravity theories, such as dCS gravity, have been proven to be well-posed \cite{Delsate:2014hba}, and third order derivatives, Ostrogradsky instability, and ghost modes can be avoided \cite{Crisostomi:2017ugk}.

Within this small-coupling approximation, many attempts have been made to construct stationary and axisymmetric BH solutions, but still, it has proven impossible to date to find exact, analytic solutions that are accurate for BHs of any spin. Two major approaches have so far been developed. 
The first approach involves assuming the BH is spinning extremely slowly, so that one can search for solutions as an additional power-series expansion in the spin parameter $a$.
In dCS gravity, this approach was initially introduced in \cite{Yunes:2009hc} to ${\cal{O}}(a)$, and then extended to ${\cal{O}}(a^2)$ in \cite{Yagi:2012ya}, to ${\cal{O}}(a^5)$ in \cite{Maselli:2015tta}, and to ${\cal{O}}(a^{15})$ in ~\cite{Cano_Ruiperez_2019}. 
In sGB gravity, this approach was initially introduced in~\cite{Ayzenberg:2014aka} to ${\cal{O}}(a^2)$, and then extended to ${\cal{O}}(a^5)$~\cite{Maselli:2015tta}, and to ${\cal{O}}(a^{15})$ in~\cite{Cano_Ruiperez_2019}. Recently, two of us have extended the dCS and sGB perturbative solution to ${\cal{O}}(a^{40})$ \cite{Chung:2024ira, Chung:2024vaf, Chung:2025gyg}. Henceforth, we shall refer to these Taylor expansions as ``series-in-$a$'' solutions. 

Although these series-in-$a$ solutions are accurate for small spins, they can \emph{never} capture the behavior of BHs with large spins due to the nature of their metric and scalar/pseudo-scalar field ansatze. 
Even for BHs with moderate spins, a lengthy expression is needed for a decently accurate representation of BH observables, as we show later in this paper. 
From an observational standpoint, this limitation is especially problematic because BH remnants produced in the collision of BH binaries observed with gravitational waves, or supermassive BHs observed with the Event Horizon Telescope, often have spin larger than one-half \cite{KAGRA:2021vkt, LIGOScientific:2025slb, Reynolds:2013rva}. 
Inaccuracies in the modeling of BHs due to the use of a slow-rotation approximation may introduce bias in tests of GR. 

To overcome the limitation of the slow-rotation approximation, another approach is to compute the metric corrections fully numerically. Such an approach has been implemented both through pseudospectral collocation methods \cite{Fernandes:2022gde, Liu:2025mfn} and finite difference methods \cite{Sullivan:2019vyi, Sullivan:2020zpf, Delgado:2020rev, Kleihaus:2011tg, Kleihaus:2015aje}. 
These methods are powerful but computationally expensive, requiring the specification of a given spin value to numerically solve for the metric and scalar/pseudo-scalar fields.
Pseudospectral methods computes spectral projections via numerical quadrature scheme, in which a large number of collocation grid points are required to obtain an accurate approximation; 
finite difference methods also require a large number of grid points to obtain accurate and smooth solutions.
This is especially important when considering multidimensional problems such as ours, where the number of grid points is a dominating factor that affects the computational cost. 

We here develop a new framework to construct analytic, closed form and accurate representations of BH solutions with arbitrary spin in a large class of beyond-Einstein effective-field theories through spectral expansions and pseudospectral collocation methods and the small-coupling approximation.
The framework begins by specifying the spacetime (and any additional fields in the theory) as a stationary and axisymmetric deformation\footnote{This assumption emerges from the small-coupling approximation and from the desire to obtain BHs that are continuously deformable to the Kerr spacetime, which implies certain solutions (like scalarized ones) cannot be captured.} of the Kerr metric in Boyer-Lindquist-like coordinates. The metric deformations and the field are then represented as a spectral series in radius and polar angle, and the coupled partial differential equations that result from the beyond-Einstein field equations are converted into a set of algebraic equations for the spectral coefficients. For any given value of black-hole spin, these algebraic equations can be readily solved for the spectral coefficients through well-established linear-algebra algorithms, such as the least-squares method or the generalized-inverse method. This procedure is then repeated on a dense grid of BH spin values, and the spectral coefficients are fitted as a function of the BH spin through non-linear functions. The result of this framework is then a fully analytic and closed-form expression for the metric and field of rotating BH in beyond-Einstein EFTs to leading-order in their coupling constants. 
The analytic expressions of the scalar field and BH spacetimes obtained in this work are contained in a \emph{Mathematica} notebook, which is available as the supplementary material of this work.

We exemplify our general framework by applying it to various specific, quadratic gravity theories: dCS, sGB, and AD gravity. In all cases, we obtain analytic, closed-form and accurate expressions for the scalar/pseudoscalar field and the BH metric with arbitrary (sub-extremal) spin. In particular, we obtain near-extremal BH solutions, whose error is below machine precision when the dimensionless spin $a \leq 0.9$. In particular, the scalar field solution is accurate to better than $10^{-12}$ when $a \leq 0.999$ and below $10^{-6}$ when $a \leq 0.9999$, while the metric is accurate to better than $10^{-12}$ when $a\leq 0.99$ and better than $10^{-8}$ when $a \leq 0.999$. 

Such analytic and accurate metrics allow for multiple applications. First, we investigate the accuracy of previous series-in-$a$ BH solutions relative to our spectral solutions. 
In particular, if we demand an error below $10^{-12}$, we find that the series-in-$a$ metrics are accurate only up to spins of $a \approx (0.2,0.6)$ depending on the accuracy sought. 
For $a > 0.6$, the series-in-$a$ solutions are inaccurate, while the metrics obtained in this work remain accurate up to the near-extremal limit. Therefore, the spectral solutions obtained here are the first analytic and accurate approximate solutions of nearly extremal BHs in beyond-Einstein theories.     

As a second application of the rotating BH metrics obtained here, we investigate physical observables, such as the surface gravity, the horizon angular velocity, the location of the innermost stable circular orbit (ISCO) and the photon ring in these theories. We find that the quadratic gravity correction to the surface gravity increases as the BH spin increases. Similarly, the dCS (sGB) correction to the angular velocity of the BH horizon is positive (negative) and increases in magnitude as the spin increases.  The dCS correction to the location of the ISCO is negative for small spins, but it flips sign at spins of about $a = 0.9$. The dCS correction to the location of the photon sphere on the equatorial plane is positive (negative) for prograde (retrograde) orbits at small spins. The distinct behavior of these quadratic-gravity corrections at large spins suggests that high-spin astrophysical observables may be ideal laboratories to detect or constrain this class of beyond Einstein effects. 

This paper explains the details of the work summarized in \cite{Lam:2025elw}. 
The rest of this paper is structured as follows. 
In Sec.~\ref{sec:QuadraticGravity}, we review the formulation of quadratic-gravity theories. 
In Sec.~\ref{sec:Method}, we lay out the details of spectral and pseudospectral methods used in obtaining the scalar field (Sec.~\ref{subsec:ScalarField}) and metric corrections (Sec.~\ref{subsec:MetricCorrections}), respectively. 
We also provide details on the numerical implementation in Sec.~\ref{subsec:NumericalImplementation}. 
In Sec.~\ref{sec:ScalarFieldResult}~and~\ref{sec:MetricCorrectionsResult}, we analyze the accuracy of the scalar field and metric corrections obtained, and compare them with the series-in-$a$ solution from \cite{Cano_Ruiperez_2019}. 
In Sec.~\ref{sec:FittedSolution}, we construct analytic fitting functions that approximate the spectral solution for all spins up to 0.999. 
These analytic solutions are then used to compute physical observables in Sec.~\ref{sec:PhysicalObservables}. 
Lastly, in Sec.~\ref{sec:Conclusion}, we summarize our results, discuss potential applications of the spectral and analytic solutions, and outline future directions.
In the remainder of the paper, we use the following conventions: $x^{\mu} = (t, r, \chi, \phi)$, where $\chi = \cos\theta$ and $\theta$ is the polar angle; the signature of the metric tensor is $(-, +, +, +)$; geometric units are used where $G = 1 = c$.  

%----------------------------------------------------------------
\section{Higher Curvature and Quadratic Gravity Theories and Perturbed Field equations}\label{sec:QuadraticGravity}

In this section, we review the class of beyond-Einstein theories we will study and we present the EFT expansions we will use to solve for spinning BH solutions. 

%------------------------------------------------
\subsection{The ABC of Higher-Curvature Effective Field Theories}
We focus, in particular, on EFTs of gravity that can be represented through a sum of higher curvature terms. The gravitational part of the Lagrangian density of these theories therefore takes the form
\begin{equation}
\begin{split}
16 \pi \mathscr{L} &= R + F_{2}(\varphi, R, R_{\mu \nu}, R_{\mu \nu \rho \sigma}, \tilde{R}_{\mu \nu \rho \sigma}) \\
&+ F_{3}(\varphi, R, R_{\mu \nu}, R_{\mu \nu \rho \sigma}, \tilde{R}_{\mu \nu \rho \sigma}) + \ldots\,,
\end{split}
\end{equation}
where $F_{A}$ is a function of scalar combinations of the Riemann curvature tensor $R_{\mu \nu \rho \sigma}$, the Ricci tensor $R_{\mu \nu}$, the Ricci scalar $R$ of $A$th order in the curvature. 
This function, in principle, can also depend on a collection of scalar, vector and tensor fields, collectively denoted $\varphi$ here. Henceforth, we shall focus on scalar-tensor theories only (so that $\varphi$ represents scalar or pseudoscalar fields only), and we shall truncate the EFT expansion at quadratic order.  We shall focus here on theories with dynamical scalar fields, and we normalize the fields so that their kinetic sectors are canonical. This allows us to rewrite the Lagrangian density as 
\begin{equation}
\begin{split}
16 \pi \mathscr{L} &= R + \sum_{q} \alpha_{q} f_{2}(\varphi_{q}) \bar{F}_{2,q}(R,R_{\mu \nu},R_{\mu \nu \rho \sigma}) \\
&- \frac{1}{2} \sum_{q} (\nabla \varphi_{q})^2 - V(\varphi_{q}) \,,
\end{split}
\end{equation}
where $q$ labels the scalar field involved, $\alpha_{q}$ is the coupling constant of the scalar field $q$, which has dimensions of length squared in geometric units, and we have further assumed a product decomposition for the function $F_{2}(\cdot)$ and we have factored-out the coupling constants $\alpha_{q}$, such that the beyond-Einstein theory reduces to GR as $\alpha_{q} \to 0$. 
In the above equation, we have used the notation $(\nabla\varphi_{q})^2 \equiv g^{\mu\nu} \nabla_{\mu}\varphi_{q} \nabla_{\nu} \varphi_{q}$ and $V(\varphi_{q})$ is a potential for the fields $\varphi_{q}$.

The method we develop here is applicable to construct spinning BHs in a wide class of beyond-Einstein theories, but to exemplify the method with a few concrete models, we shall further specialize our calculations to the following Lagrangian density \cite{Cano_Ruiperez_2019}:
\begin{equation}
\begin{split}
    16 \pi \mathscr{L} &= R - \frac{1}{2} (\nabla \varphi_{\rm sGB})^2 - \frac{1}{2} (\nabla \varphi_{\rm dCS})^2 - V(\varphi_{\rm sGB}, \varphi_{\rm dCS}) \\
    &+ \alpha_{\rm sGB} \, f_{2,\rm sGB}(\varphi_{\rm sGB}, \varphi_{\rm dCS}) \mathscr{Q}_{\rm sGB} \\
    &+ \alpha_{\rm dCS} \, f_{2,\rm dCS}(\varphi_{\rm sGB}, \varphi_{\rm dCS}) \mathscr{Q}_{\rm dCS}, 
    \label{eq:Lag-fam}
\end{split}
\end{equation}
where $\varphi_{\rm sGB}$ and $\varphi_{\rm dCS}$ are dynamical scalar and pseudoscalar fields respectively, which we collectively refer to as the ``scalar fields.'' In this specialization, we have further assumed diagonal kinetic terms for each of the fields, and specific non-minimal couplings, namely the Gauss-Bonnet invariant $\bar{F}_{2,\rm sGB} = \mathscr{Q}_{\rm sGB}$ and the Pontryagin invariant $\bar{F}_{2,\rm dCS} = \mathscr{Q}_{\rm dCS}$, where
\begin{equation}
\begin{split}
    \mathscr{Q}_{\rm sGB} &= R_{\mu \nu \rho \sigma} R^{\mu \nu \rho \sigma} - 4 R_{\mu \nu} R^{\mu \nu} + R^2, \\
    \mathscr{Q}_{\rm dCS} &= R_{\mu \nu \rho \sigma} \tilde{R}^{\mu \nu \rho \sigma}, 
\end{split}
\end{equation}
with the dual Riemann tensor defined through
\begin{equation}
\tilde{R}^{\mu \nu \rho \sigma} = \frac{1}{2} \epsilon^{\rho \sigma \alpha \beta} R^{\mu \nu} {}_{\alpha \beta}. 
\end{equation}
and the Levi-Civita tensor defined through 
\begin{equation}
\epsilon^{\rho \sigma \alpha \beta} = \frac{1}{\sqrt{-g}} [\rho \sigma \alpha \beta], 
\end{equation}
where $g$ is the determinant of $g_{\mu \nu}$, and $[\rho \sigma \alpha \beta]$ is the totally antisymmetric Levi-Civita symbol, with convention $[t r \chi \phi] = 1$~\cite{Chung:2025gyg}. The coupling constants $\alpha_{\rm sGB, dCS}$ have dimensions of length squared in geometric units, given our choices for the kinetic term. 

The Lagrangian in Eq.~\eqref{eq:Lag-fam} still defines a family of theories of gravity, specific by the potential $V(\varphi_{\rm sGB}, \varphi_{\rm dCS})$ and the choice of coupling functions $f_{2,\rm sGB}$ and  $f_{2,\rm dCS}$. In order to concretely exemplify our general method, we must therefore eliminate this additional freedom to select specific members in this family of beyond-Einstein theories.  We will here consider massless scalar fields with shift-symmetric coupling functions \cite{Cano_Ruiperez_2019, Chung:2024vaf, Chung:2024ira, Chung:2025gyg}, i.e., $V = 0$, while $f_{2,\rm sGB,dCS}$ are assumed linear in $\varphi_{\rm sGB,dCS}$. 
The shift-symmetric assumption can be regarded as a leading-order approximation in small coupling $|\varphi_{\rm sGB, dCS}| \ll 1$ of a more sophisticated coupling function, because $\mathscr{Q}_{\rm sGB, dCS}$ are topological invariants \cite{Chung:2024vaf}. 
We shall adopt the mixing-angle representation of these coupling functions, namely \cite{Cano_Ruiperez_2019}
\begin{equation}
\begin{split}
    f_1(\varphi_{\rm sGB}, \varphi_{\rm dCS}) &= \varphi_{\rm sGB} \\
    f_2(\varphi_{\rm sGB}, \varphi_{\rm dCS}) &= \sin\theta_m \varphi_{\rm sGB} + \cos\theta_m \varphi_{\rm dCS}, 
\end{split}
\end{equation}
where $\theta_m$ is a mixing angle that characterizes the amount of parity mixing between $\varphi_{\rm sGB}$ and $\varphi_{\rm dCS}$. 

The above choices lead to 3 distinct classes of theories that we will apply our framework to as direct examples.  
The first example is sGB gravity~\cite{Yagi:2015oca, Maeda:2009uy, Moura:2006pz}, which can be obtained by setting $\alpha_{\rm dCS} = 0$. In this case, $\varphi_{\rm dCS}$ decouples, leaving only the kinetic term $\frac{1}{2}(\nabla \varphi_{\rm dCS})^2$ in the Lagrangian density, which leads to a vanishing $\varphi_{\rm dCS}$ after imposing appropriate boundary conditions\footnote{This kinetic term leads to a wave equation for $\varphi_{\rm dCS}$ that is unsourced. Therefore, once initial data is chosen, this wave propagates out to future null infinity at the speed of light, leaving the computational domain for good. }. 
The second example is dCS gravity~\cite{Jackiw:2003pm, Yunes:2009hc} , which can be obtained by setting $\alpha_{\rm sGB} = 0 = \theta_m$. In this case, $\varphi_{\rm sGB}$ decouples just as $\varphi_{\rm dCS}$ did in the previous case. 
The third example is AD gravity  \cite{Cano:2021rey, Kanti:1995cp}, which can be obtained by setting $\alpha_{\rm AD} \equiv \alpha_{\rm sGB} = \alpha_{\rm dCS}$ and $\theta_m = 0$.
In this case, since the AD Lagrangian density is the sum of the sGB and dCS Lagrangian densities, the corresponding dilaton (axion) field is $\varphi_{\rm sGB}$ ($\varphi_{\rm dCS}$) and the metric corrections are just the direct sum of the sGB and dCS contributions (up to higher powers of $\alpha_{\rm sGB,dCS}$ \cite{Chung:2025wbg}). 
When exemplifying our method, therefore, we will focus on the sGB and dCS cases for the most part. 

Before proceeding, let us write the Lagrangian density for shift-symmetric massless sGB and dCS gravity more compactly as
\begin{equation}
    16 \pi \mathscr{L}_q = R - \frac{1}{2} (\nabla \varphi_q)^2 + \alpha_q \varphi_q \mathscr{Q}_q, 
\end{equation}
where $q = \rm sGB, dCS$.  
Varying the action with respect to the metric and scalar fields, we obtain the following vacuum field equations:
\begin{align}
    R_{\mu}{}^{\nu} + \zeta_q [(\mathscr{A}_q)_{\mu}{}^{\nu} - (\bar{T}_q)_{\mu}{}^{\nu}] &= 0, \label{eq:MEE} \\
    \Box \vartheta_q + \mathscr{Q}_q &= 0, \label{eq:SFE}
\end{align}
where $\zeta_q = \alpha_q^2/M^4$ is a dimensionless coupling parameter, $M$ is the ADM mass of the BH\footnote{Strictly speaking, $\zeta_q = \alpha_q^2/\lambdabar^4$, where $\lambdabar$ is a characteristic length scale. In our work, a natural length scale is the ADM mass of the BH.}, and $\Box = \nabla_{\mu} \nabla^{\mu}$ is the d'Alembertian operator, with $\vartheta_q = \varphi_q/\alpha_q$ a rescaled scalar field. 
The quantity $(\mathscr{A}_q)_{\mu}{}^{\nu}$ is a (1, 1) tensor constructed from a product of the curvature tensor and derivatives of the scalar fields.
For sGB gravity, $(\mathscr{A}_{\rm sGB})_{\mu}{}^{\nu}$ is explicitly given by \cite{East:2020hgw, Cano_Ruiperez_2019} 
\begin{equation}\label{eq:AuusGB}
\begin{split}
    (\mathscr{A}_{\rm sGB})_{\mu}{}^{\nu} &= \left[\delta_{\mu \lambda \gamma \delta}^{\nu \sigma \alpha \beta} - \frac{1}{2} \delta_{\mu}{}^{\nu}\delta_{\eta \lambda \gamma \delta}^{\eta \sigma \alpha \beta}\right] R^{\gamma \delta}{}_{\alpha \beta} \nabla^{\lambda} \nabla_{\sigma} \vartheta_{\rm sGB}, 
\end{split}
\end{equation}
where $\delta^{\nu \sigma \alpha \beta}_{\mu \lambda \gamma \delta}$ is the generalized Kronecker delta, defined as 
\begin{equation}
\delta_{\mu_1 \mu_2 \mu_3 \mu_4}^{\nu_1 \nu_2 \nu_3 \nu_4} = \det 
\begin{pmatrix}
\delta_{\mu_1}^{\nu_1} & \delta_{\mu_2}^{\nu_1} & \delta_{\mu_3}^{\nu_1} & \delta_{\mu_4}^{\nu_1} \\
\delta_{\mu_1}^{\nu_2} & \delta_{\mu_2}^{\nu_2} & \delta_{\mu_3}^{\nu_2} & \delta_{\mu_4}^{\nu_2} \\
\delta_{\mu_1}^{\nu_3} & \delta_{\mu_2}^{\nu_3} & \delta_{\mu_3}^{\nu_3} & \delta_{\mu_4}^{\nu_3} \\
\delta_{\mu_1}^{\nu_4} & \delta_{\mu_2}^{\nu_4} & \delta_{\mu_3}^{\nu_4} & \delta_{\mu_4}^{\nu_4} \\
\end{pmatrix}. 
\end{equation}
For dCS gravity, $(\mathscr{A}_{\rm dCS})_{\mu}{}^{\nu} = g_{\mu \alpha} (\mathscr{A}_{\rm dCS})^{\alpha \nu}$, where $(\mathscr{A}_{\rm dCS})^{\mu \nu}$ is explicitly given by \cite{Yunes:2009hc}
\begin{equation}\label{eq:AuudCS}
\begin{split}
    (\mathscr{A}_{\rm dCS})^{\mu\nu} &= -4\Big[\left(\nabla_\sigma \vartheta_{\rm dCS} \right) \varepsilon^{\sigma \delta \alpha(\mu|} \nabla_\alpha R^{|\nu)}{}_{\delta} \\
    &\qquad\qquad + \left(\nabla_\sigma \nabla_\delta \vartheta_{\rm dCS} \right) \tilde{R}^{\delta (\mu \nu) \sigma}\Big].
\end{split}
\end{equation}
The trace-reversed stress-energy tensor is given by
\begin{equation}
    (\bar{T}_q)_{\mu}{}^{\nu} = \frac{1}{2}\nabla_{\mu} \vartheta_q \nabla^{\nu} \vartheta_q. 
\end{equation}

%----------------------------------------
\subsection{Perturbed Effective-Field-Theory-Corrected Field Equations}
We begin by perturbatively expanding the metric and scalar field in powers of $\zeta$ as follows:
\begin{align}
    g_{\mu\nu} &= g_{\mu\nu}^{(0)} + \zeta g_{\mu\nu}^{(1)} + {\cal{O}}(\zeta^2), \\
    \vartheta &= \vartheta^{(0)} + {\cal{O}}(\zeta), 
\end{align}
where the superscript ${(k)}$ on $g_{\mu\nu}$ and $\vartheta$ represents quantities of ${\cal{O}}(\zeta^k)$. 
For clarity's sake, we drop the subscript $q$ that indicates sGB or dCS gravity below.

Substituting the expansion into Eqs.~(\ref{eq:MEE}) and (\ref{eq:SFE}), the zeroth-order-in-$\zeta$ equations can be written schematically as 
\begin{align}
    [R_{\mu}{}^{\nu}]^{(0)} &= 0, \label{eq:EE0} \\
    E_{\vartheta} \equiv \Box^{(0)} \vartheta + \mathscr{Q}^{(0)} &= 0. \label{eq:SF0}
\end{align}
Notice that we have dropped the superscript zero on $\vartheta$, as we are only going to compute $\vartheta$ to leading order, $\vartheta = \vartheta^{(0)}$. 
Equation~(\ref{eq:EE0}) is the standard Einstein field equation in vacuum, which has the Kerr metric as its vacuum, stationary, axisymmetric solution. We adopt coordinates $(t,r,\chi,\phi)$ such that the Kerr metric takes the following form:
\begin{equation}
\begin{split}
g_{\mu \nu}^{(0)} dx^{\mu}  dx^{\nu} &= - \left( 1-\frac{2 M r}{\Sigma}\right) dt^2 - \frac{4 M^2 a r}{\Sigma} (1 - \chi^2) d \phi dt \\
& \quad + \left( \frac{\Sigma}{\Delta} dr^2 + \frac{\Sigma}{1 - \chi^2} d \chi^2 \right) \\
& \quad + (1-\chi^2) \\
& \quad \quad \times \left[r^{2} + M^2 a^{2}+\frac{2 M^3 a^{2} r}{\Sigma} (1 - \chi^2)\right] d\phi^2,
\end{split}
\end{equation}
where $\Sigma = r^2 + M^2 a^2 \chi^2$, $\Delta = (r - r_+)(r - r_-)$, and $r_{\pm} = M(1 \pm \sqrt{1-a^2})$. 

Equation~(\ref{eq:SF0}) is the evolution equation for the scalar field at leading order in $\zeta$, which governs the dynamical behavior of $\vartheta$. 
Note that $\Box^{(0)}$ and $\mathscr{Q}^{(0)}$ are evaluated with respect to the Kerr metric. Explicitly, Eq.~(\ref{eq:SF0}) can be written using 
\begin{equation}\label{eq:SFexplicit}
\begin{split}
    \Box^{(0)} \vartheta &= \frac{1}{\Sigma} \left[\pdv{r}\left(\Delta \pdv{\vartheta}{r}\right) + \pdv{\chi}\left((1-\chi^2)\pdv{\vartheta}{\chi}\right)\right], \\
    \mathscr{Q}_{\rm sGB}^{(0)} &= \frac{48(r^6 - 15M^2a^2r^4\chi^2 + 15M^4a^4r^2\chi^4 - M^6a^6\chi^6)}{\Sigma^6}, \\
    \mathscr{Q}_{\rm dCS}^{(0)} &= -\frac{96 M^3 a r \chi (r^2-3M^2a^2\chi^2)(3r^2-M^2a^2\chi^2)}{\Sigma^6}.
\end{split}
\end{equation}
in sGB and dCS gravity.  Thus, Eqs.~\eqref{eq:SF0} and \eqref{eq:SFexplicit} define a linear partial differential equation for the field $\vartheta$ that needs to be solved. 

To obtain the leading-order-in-$\zeta$ correction to the metric, induced by the back-reaction of the scalar field onto the metric, we peel out the following asymptotic behavior for the metric correction $g_{\mu \nu}^{(1)}$~\cite{Cano_Ruiperez_2019}:
\begin{equation}\label{eq:metric}
\begin{split}
g_{\mu \nu}^{(1)} dx^{\mu}  dx^{\nu} & = H_1(r, \chi) dt^2 \\
& \quad - H_2(r, \chi) \frac{4 M^2 a r}{\Sigma} (1 - \chi^2) d \phi dt \\
& \quad + H_3(r, \chi) \left( \frac{\Sigma}{\Delta} dr^2 + \frac{\Sigma}{1 - \chi^2} d \chi^2 \right) \\
& \quad + H_4(r, \chi) (1-\chi^2) \\
& \quad \quad \times \left[r^{2} + M^2 a^{2}+\frac{2 M^3 a^{2} r}{\Sigma} (1 - \chi^2)\right] d\phi^2,
\end{split}
\end{equation}
where $H_{i}$ are functions of $r$ and $\chi$ only, and independent of $\zeta$. 
This metric ansatz has the advantage that the horizon is specified by the roots of $\Delta = 0$, regardless of the modifications $H_i(r,\chi)$, which, in turn, implies that the outer horizon remains at $r = r_+$. 
Furthermore, the metric correction enforces circularity, which is guaranteed in EFT-corrected black holes \cite{Xie:2021bur}. 

We also impose peeling conditions on the $H_i$ functions so that $M$ and $J = M^2 a$ remain the ADM mass and the ADM angular momentum of the EFT-corrected black hole. 
More specifically, we require that 
\begin{equation}\label{eq:BoundaryCondition}
    H_1^{(0)} = 0, \quad H_2^{(0)} = \frac{H_3^{(1)}}{2M}, \quad H_3^{(0)} = H_4^{(0)} = -\frac{H_3^{(1)}}{M}, 
\end{equation}
where the functions $H_i^{(0)}(\chi)$ and $H_i^{(1)}(\chi)$ are defined by 
\begin{align} 
H_i(r, \chi) = H_i^{(0)}(\chi) + \frac{1}{r} H_i^{(1)}(\chi) + {\cal{O}}(r^{-2})\,,
\end{align}
With these conditions, $M$ and $a$ can still be interpreted as the mass and dimensionless spin of the EFT-corrected black hole \cite{Cano_Ruiperez_2019}. 

To compute $H_i$, we perturbatively expand Eq.~(\ref{eq:MEE}) to first order in $\zeta$ to find
\begin{align}\label{eq:EE1}
    E_{\mu}{}^{\nu} \equiv [R_{\mu}{}^{\nu}]^{(1)} + [\mathscr{A}_{\mu}{}^{\nu}]^{(0)} - [\bar{T}_{\mu}{}^{\nu}]^{(0)} = 0. 
\end{align}
Inserting the metric ansatz, in general, $[R_{\mu}{}^{\nu}]^{(1)}$ contains a linear combination of $H_i$ and their derivatives, with rational functions of $r$ and $\chi$ as coefficients. 
Schematically, we write $[R_{\mu}{}^{\nu}]^{(1)} = (\mathscr{D}_{i})_{\mu}{}^{\nu} H_i$, where $(\mathscr{D}_{i})_{\mu}{}^{\nu}$ is a linear differential operator with respect to $r$ and $\chi$, and $i$ is being summed from 1 to 4. Thus, Eq.~\eqref{eq:EE1} is a linear partial differential equation for the $H_i$ functions that needs to be solved. 

%----------------------------------------------------------------
\section{Spectral and pseudospectral collocation methods}\label{sec:Method}

In this section, we employ pseudospectral methods and spectral expansions to find analytic, closed-form solutions for spinning BHs (with any spin) in a wide class of EFT-corrected gravity theories, including sGB gravity, dCS gravity and axidilaton gravity, to leading order in the small-coupling approximation.

Let us then begin by briefly reviewing (pseudo)spectral methods. The reader who is interested in more details can refer to \cite{boyd2013chebyshev}, while those already versed in these methods can skip to the next subsection. (Pseudo)spectral methods start by assuming the solutions are linear combinations of complete, orthogonal functions, e.g., Chebyshev or Legendre polynomials. 
For spectral methods, one projects the metric corrections and field equations onto individual basis functions, yielding a set of \emph{algebraic} equations that can be solved using common matrix algebra techniques. 
For pseudospectral methods, instead of computing the projection analytically, one instead uses quadratures as approximations. 
This, in practice, is equivalent to evaluating the source on some grid points, such as Chebyshev ``roots'' or Legendre-Lobatto grid points. 
Thus, pseudospectral methods avoid evaluating complicated integrals that involve basis functions, though the rate of convergence is often less optimal than in the spectral method case \cite{boyd2013chebyshev}. 

In general, there are multiple advantages to using spectral and pseudospectral methods over finite difference methods, which we summarize as follows: 
\begin{enumerate}
    \item Instead of having an algebraic rate of convergence, exponential convergence is guaranteed \cite{boyd2013chebyshev}. This reduces the spectral order (number of basis functions) required, implying better accuracy and lower memory requirements;
    \item The desired boundary conditions can be better prescribed through a customized \emph{ansatz} and a suitable choice of basis functions tailored to the problem. 
    In particular, one can (i) choose an ansatz that satisfies boundary conditions automatically (behavioral boundary conditions), and (ii) choose basis functions that satisfy the symmetries of the problem;
    \item Unlike finite difference methods, spectral solutions, as linear combinations of basis functions, are automatically smooth. Thus, no interpolation or fitting between grid points is needed. 
\end{enumerate}

In the context of GR, (pseudo)spectral methods were first used in constructing BH initial data, and applied to BH simulations intending to obtain smooth solutions \cite{Kidder:1999fv, Kidder:2000yq, Pfeiffer:2003amu}. 
These methods were later applied to various areas, including asteroseismology \cite{Ferrari:2007rc, Gaertig:2008uz}, BH quasinormal modes \cite{Langlois:2021xzq, Langlois:2021aji, Chung:2023zdq, Chung:2023wkd, Chung:2024vaf, Chung:2024ira, Chung:2025gyg, Blazquez-Salcedo:2023hwg, Blazquez-Salcedo:2024oek}, and more recently, to constructing beyond-Einstein BH solutions \cite{Dias:2015nua, Fernandes:2022gde, Fernandes:2024ztk, Liu:2025mfn}. 
Since exterior BH solutions are smooth in nature, these methods are especially adapted to solving the field equations. 
In particular, \cite{Fernandes:2022gde, Liu:2025mfn} applied pseudospectral methods to find BH solutions in sGB gravity without invoking the small-coupling approximation and slow-rotation expansion. 
These studies showcase the possibility of computing BH solutions in beyond-Einstein gravity. 

%---------------
\subsection{Scalar field}\label{subsec:ScalarField}

Let us now solve Eq.~\eqref{eq:SF0} using pseudospectral methods, which using Eq.~\eqref{eq:SFexplicit} becomes
\begin{align}
\pdv{r}\left(\Delta \pdv{\vartheta}{r}\right) + \pdv{\chi}\left[(1-\chi^2)\pdv{\vartheta}{\chi}\right]
&=  -\Sigma \, \mathscr{Q}_{q}^{(0)}\,,\label{eq:SF0-final}
\end{align}
where $\mathscr{Q}_{\rm sGB/dCS}^{(0)}$ is given in Eq.~\eqref{eq:SFexplicit}. Motivated by the Legendre differential operator in $\Box^{(0)}$, we decompose $\vartheta$ into a sum of Legendre modes $P_{\ell}(\chi)$, i.e., 
\begin{equation}
    \vartheta(r,\chi) = \sum_{\ell=0}^{\infty} \vartheta_{\ell}(r) P_{\ell}(\chi), \label{eq:varthetaLegendreExpansion} 
\end{equation}
and Eq.~(\ref{eq:SF0-final}) becomes
\begin{equation}\label{eq:vartheta_ell_Eq}
    \mathcal{D}\vartheta_{\ell} \equiv \dv{r}\left(\Delta \dv{\vartheta_{\ell}}{r}\right) - \ell(\ell+1) \vartheta_{\ell} = s_{\ell}(r), 
\end{equation}
where 
\begin{equation}
    s_{\ell}(r) = -\frac{2\ell + 1}{2} \int_{-1}^{+1} \Sigma \mathscr{Q}^{(0)} P_{\ell}(\chi) \, d\chi. 
\end{equation}
In practice, it is impossible to incorporate infinitely many Legendre modes in Eq.~\eqref{eq:varthetaLegendreExpansion}. 
We truncate this expansion at some finite spectral order $N_{\chi}$. 

When solving Eq.~(\ref{eq:vartheta_ell_Eq}), $\vartheta_{\ell}(r)$ can be split into a homogeneous and a particular solution. The homogeneous solutions do not satisfy the boundary conditions (regularity at the horizon or asymptotic flatness at spatial infinity), and thus, it must be discarded.
We can find the particular solution by first solving analytically for $s_{\ell}(r)$, and one finds \cite{McNees:2015srl, Stein:2014wza}
\begin{widetext}
\begin{equation}\label{eq:ScalarFieldSource}
\begin{split}
    s_{\ell, \rm sGB}(r) &= 4 \Re \bigg\{ (-1)^{\frac{\ell}{2}} \frac{\Gamma(\frac{1}{2})\Gamma(\ell+4)}{2^{\ell}\Gamma(\ell+\frac{1}{2})}\frac{a^{\ell}}{r^{\ell+4}} \bigg[3\,{}_2F_1\left(\frac{\ell+4}{2},\frac{\ell+5}{2};\ell+\frac{3}{2};-\frac{a^2}{r^2}\right) - (\ell+5)\,{}_2F_1\left(\frac{\ell+4}{2},\frac{\ell+7}{2};\ell+\frac{3}{2};-\frac{a^2}{r^2}\right)\bigg]\bigg\}, \\
    s_{\ell, \rm dCS}(r) &= 4 \Im \bigg\{ (-1)^{\frac{\ell}{2}} \frac{\Gamma(\frac{1}{2})\Gamma(\ell+4)}{2^{\ell}\Gamma(\ell+\frac{1}{2})}\frac{a^{\ell}}{r^{\ell+4}} \bigg[3\,{}_2F_1\left(\frac{\ell+4}{2},\frac{\ell+5}{2};\ell+\frac{3}{2};-\frac{a^2}{r^2}\right) - (\ell+5)\,{}_2F_1\left(\frac{\ell+4}{2},\frac{\ell+7}{2};\ell+\frac{3}{2};-\frac{a^2}{r^2}\right)\bigg]\bigg\}. 
\end{split}
\end{equation}
\end{widetext}
Notice $s_{\ell, \rm sGB}(r) = 0$ for odd $\ell$, while $s_{\ell, \rm dCS}(r) = 0$ for even $\ell$. Hence, only the even (odd) Legendre polynomials will enter Eq.~(\ref{eq:varthetaLegendreExpansion}) in sGB (dCS) gravity. 

With this symmetry, we first write Eq.~(\ref{eq:varthetaLegendreExpansion}) as 
\begin{equation}\label{eq:varthetaParityLegendreExpansion} 
\begin{split}
    \vartheta_{\rm sGB}(r,\chi) &= \sum_{\mathrm{even} \, \ell}^{N_{\chi}} \vartheta_{\ell, \rm sGB}(r) P_{\ell}(\chi) \\
    \vartheta_{\rm dCS}(r,\chi) &= \sum_{\mathrm{odd} \, \ell}^{N_{\chi}} \vartheta_{\ell, \rm dCS}(r) P_{\ell}(\chi)
\end{split}
\end{equation}
and we further incorporate the asymptotic behavior of the particular solution at spatial infinity, which improves the accuracy of our spectral solution, as follows.
Using the method of variation of parameters, one can find that $\vartheta_{\ell} \sim r^{-\ell-1}$ at spatial infinity~\cite{McNees:2015srl}. 
We thus define $\tilde{\vartheta}_{\ell}$ such that $\vartheta_{\ell} = r^{-\ell-1} \tilde{\vartheta}_{\ell}$, and then Eq.~(\ref{eq:vartheta_ell_Eq}) becomes
\begin{equation}\label{eq:vartheta_tilde_ell_Eq}
    \tilde{\mathcal{D}}_{\ell} \tilde{\vartheta}_{\ell} \equiv \dv{r}\left(\Delta\dv{r}r^{-\ell-1}\tilde{\vartheta}_{\ell}\right) - \ell(\ell+1)r^{-\ell-1}\tilde{\vartheta}_{\ell} = s_{\ell}(r)\,,
\end{equation}
which is just an ordinary differential equation in radius with a non-trivial source.

We follow~\cite{Stein:2014xba} and solve this differential equation using the pseudospectral collocation method. 
First, we compactify\footnote{Our compactified coordinate is different from that used in~\cite{Stein:2014xba}, where the author used $z_{\rm other} = 1 - 2(r_+ - 1)/(r - 1)$ to map $(r_+, \infty)$ to $(-1,+1)$.} the radial domain using \cite{Langlois:2021xzq,Jansen:2017oag}
\begin{align}\label{eq:compactify}
    z = \frac{2r_+}{r} - 1, 
\end{align}
which maps spatial infinity to $z=-1$ and the BH event horizon to $z=1$. 
We then solve Eq.~(\ref{eq:vartheta_tilde_ell_Eq}) by writing $\tilde{\vartheta}_{\ell}$ as a linear combination of Chebyshev polynomials in $z$, 
\begin{align}\label{eq:vartheta_ell_expansion}
    \tilde{\vartheta}_{\ell}(z) = \sum_{n=0}^{N_{z}} c_{n\ell} T_n(z),
\end{align}
where $N_{z}$ is the spectral order that we truncate at. 
Substituting Eq.~(\ref{eq:vartheta_ell_expansion}) into Eq.~(\ref{eq:vartheta_tilde_ell_Eq}), we find 
\begin{equation}\label{eq:vartheta_tilde_ell_Eq_compactified}
    \sum_{n=0}^{N_z} c_{n\ell} \hat{\tilde{\mathcal{D}}}_{\ell} T_n(z) = s_{\ell}(z). 
\end{equation}
where
\begin{equation}
\begin{split}
    \hat{\tilde{\mathcal{D}}}_{\ell} T_n(z) &\equiv \frac{(1+z)^2}{2r_+} \dv{z} \left( \Delta \frac{(1+z)^2}{2r_+} \dv{z} \left[\left( \frac{2r_+}{1+z} \right)^{-\ell-1} T_n(z)\right]\right) \\
    & \quad - \ell(\ell+1) \left( \frac{2r_+}{1+z} \right)^{-\ell-1} T_n(z).
\end{split}
\end{equation}
If we were to attempt to solve for the $c_{n\ell}$ coefficients using a spectral method, we would have to project both sides of this equation to the Chebyshev basis. 
For $s_{\ell}(z)$, the projection amounts to integrating a product of hypergeometric functions in Eqs.~(\ref{eq:ScalarFieldSource}) and $T_{n'}(z) (1-z^2)^{-1/2}$. 
However, we were unable to find an analytic solution to such integrals, and thus, we will instead employ a pseudospectral collocation method below. 

In pseudospectral collocation methods, one starts by evaluating the ordinary differential equations on the Chebyshev ``roots'' or grid points 
\begin{equation}
z_k = -\cos\left[\frac{(2k+1)\pi}{2(N+1)}\right], 
\end{equation}
where $k = 0,1,\ldots,N_z$. Doing so reduces Eq.~(\ref{eq:vartheta_tilde_ell_Eq_compactified}) into the matrix equation
\begin{equation}
    \sum_{n=0}^{N_z} c_{n\ell} [\hat{\tilde{\mathcal{D}}}_{\ell} T_n(z_k)] = s_{\ell}(z_k), \quad \forall k=0,1,\ldots,N_z. 
\end{equation}
Written more compactly, we have
\begin{equation}\label{eq:PseudospectralMatEq}
    \mathfrak{D}_{\ell} \mathbf{c}_{\ell} = \mathfrak{s}_{\ell}, 
\end{equation}
where the coefficient vector $\mathbf{c}_{\ell} = (c_{0 \ell},c_{1 \ell},\ldots,c_{N_z \ell})$ and source vector $\mathfrak{s}_{\ell} = (s_{\ell}(z_0), s_{\ell}(z_1), \ldots, s_{\ell}(z_{N_z}))$ are $(N_z+1)$-vectors, and $[\mathfrak{D}_{\ell}]_{kn} = [\hat{\tilde{\mathcal{D}}}_{\ell} T_n(z_k)]$ is a $(N_z+1)\times(N_z+1)$ matrix (with rows taking different values of $z_k$ and columns representing different values of $n$ in $T_n$).
Since we have already peeled off the asymptotic behavior of the scalar field at spatial infinity and we are considering particular solutions, no explicit boundary conditions have to be imposed on $\tilde{\theta}_{\ell}$. 
Thus, inverting Eq.~(\ref{eq:PseudospectralMatEq}), $\mathbf{c}_{\ell} = \mathfrak{D}_{\ell}^{-1} \mathfrak{s}_{\ell}$, yields the scalar field solution in spectral form. 
The scalar field solution can then be reconstructed using Eq.~(\ref{eq:vartheta_ell_expansion}) and (\ref{eq:varthetaLegendreExpansion}). We will explore the accuracy of this solution as a function of spectral order and BH spin in the next section.

\subsection{Metric corrections}\label{subsec:MetricCorrections}

Let us now solve Eq.~\eqref{eq:EE1} using spectral expansions. 
The tensors $[\mathscr{A}_{\mu}{}^{\nu}]^{(0)}$ and $[\bar{T}_{\mu}{}^{\nu}]^{(0)}$ do not contain any $H_i$, as they are zeroth order quantities\footnote{
In the case of dCS gravity, we can simplify $[(\mathscr{A}_{\rm dCS})_{\mu}{}^{\nu}]^{(0)}$ using the fact that it is a zeroth-order quantity. 
Since Eq.~(\ref{eq:AuudCS}) is computed with respect to the Kerr background, its first term vanishes, and one is left only with \cite{Li:2022pcy, Chung:2025gyg}
\begin{align}
    [(\mathscr{A}_{\rm dCS})_{\mu}{}^{\nu}]^{(0)} = -4 (\nabla_{\sigma} \nabla_{\delta} \vartheta_{\rm dCS}) \tilde{R}^{\delta (\mu \nu) \sigma}. 
\end{align}
}.
Therefore, with the scalar field in hand from the previous subsection, $[\mathscr{A}_{\mu}{}^{\nu}]^{(0)}$ and $[T_{\mu}{}^{\nu}]^{(0)}$ now act as source terms, and Eq.~(\ref{eq:EE1}) can be cast into
\begin{equation}\label{eq:EE1Schematic}
    \sum_{i = 1}^{4} \ (\mathscr{D}_{i})_{\mu}{}^{\nu} H_i = - [\mathscr{A}_{\mu}{}^{\nu}]^{(0)} + [\bar{T}_{\mu}{}^{\nu}]^{(0)} \equiv S_{\mu}{}^{\nu}
\end{equation}
Observe that $S_{\mu}{}^{\nu}$ contains only information about the scalar field and the Kerr metric, which sources $H_i$ through a particular solution. The general solution to Eq.~\eqref{eq:EE1Schematic} will then be the sum of this particular solution and a homogeneous solution, chosen by the boundary conditions stated in Eq.~(\ref{eq:BoundaryCondition}). 

For each component of Eq.~(\ref{eq:EE1Schematic}), we multiply through the common denominator and obtain
\begin{equation}\label{eq:EE1r}
    \sum_{i=1}^{4} \sum_{\delta,\sigma} \sum_{\alpha,\beta=0}^{2} \mathcal{G}_{i,\delta,\sigma,\alpha,\beta}^{j} r^{\delta} \chi^{\sigma} \partial_{r}^{\alpha} \partial_{\chi}^{\beta} H_i(r,\chi) = \sum_{\delta,\sigma} \mathcal{S}_{\delta,\sigma}^{j} r^{\delta} \chi^{\sigma}, 
\end{equation}
where $\mathcal{G}_{i,\delta,\sigma,\alpha,\beta}^{j}$ are coefficients that depend on $M$ and $a$, and $\mathcal{S}_{\delta,\sigma}^{j}$ are coefficients that depend on $c_{n\ell}$, $M$ and $a$, and $j$ represents the number of equations that has to be solved.
In general, Eq.~\eqref{eq:EE1r} contains ten independent equations, but we find that the $(\mu,\nu) = (t,r), (t,\chi), (r,\phi), (\chi,\phi)$ components are identically zero. 
Out of the remaining 6 equations, we choose to solve the $(t,t)$, $(t,\phi)$, $(r,r)$ and $(\chi,\chi)$ components ($j=1,2,3,4$), as there are only four $H_i$ functions, as in~\cite{Cano_Ruiperez_2019}.
In Sec.~\ref{sec:MetricCorrectionsResult}, we check that all six equations are actually satisfied, and it suffices to only consider the four that we have chosen above. 

We now use Eq.~(\ref{eq:compactify}) to compactify the radial coordinate in Eq.~(\ref{eq:EE1r}) and we multiply it by its common denominator, giving us 
\begin{equation}\label{eq:EE1z}
    \sum_{i=1}^{4} \sum_{\delta,\sigma} \sum_{\alpha,\beta=0}^{2} \tilde{\mathcal{G}}_{i,\delta,\sigma,\alpha,\beta}^{j} z^{\delta} \chi^{\sigma} \partial_{z}^{\alpha} \partial_{\chi}^{\beta} H_i(z,\chi) = \sum_{\delta,\sigma} \tilde{\mathcal{S}}_{\delta,\sigma}^{j} z^{\delta} \chi^{\sigma}, 
\end{equation}
where $\tilde{\mathcal{G}}_{i,\delta,\sigma,\alpha,\beta}^{j}$ are coefficients that depend on $M$ and $a$, and $\tilde{\mathcal{S}}_{\delta,\sigma}^{j}$ are coefficients that depend on the spectral expansion coefficients of the scalar field $c_{n\ell}$, as well as $M$ and $a$. 
As shown in~\cite{Cano:2022wwo, Cano_Ruiperez_2019} when considering small-spin expansions, representing $H_i$ as powers of $1/r$ and $\chi$ yields the correct series-in-$a$ solutions.
We here wish to generalize this work to moderate and large spins, so we instead write the metric corrections as a sum of Chebyshev polynomials in $z$ and Legendre polynomials in $\chi$, namely
\begin{equation}\label{eq:HAnsatz}
    H_i(z,\chi) = \sum_{n=0}^{\mathcal{N}_z} \sum_{\ell=0}^{\mathcal{N}_{\chi}} v_{n\ell}^{i} T_{n}(z) P_{\ell}(\chi), 
\end{equation}
where $\mathcal{N}_z$ and $\mathcal{N}_{\chi}$ are the spectral order of the series. Chebyshev and Legendre polynomials satisfy certain orthonormality conditions, and also lead to exponential convergence, as we will show in Sec.~\ref{sec:MetricCorrectionsResult}.

Observe that the source term $S_{\mu}{}^{\nu}$ for both sGB and dCS theory has reflection symmetry, i.e., it is even in $\chi$. 
For these two theories, this implies the particular solution computed would also be even in $\chi$. 
We further assume the homogeneous solutions determined by the boundary conditions [Eq.~(\ref{eq:BoundaryCondition})] also satisfy reflection symmetry. 
Physically, it is reasonable that including a source term that is even in $\chi$ would not change the parity of the metric solution. 
Therefore, by imposing $v_{n\ell}^i = 0$ for odd $\ell$, we drop the odd terms in Eq.~(\ref{eq:HAnsatz}), which reduces the number of variables by half. 

Substituting Eq.~(\ref{eq:HAnsatz}) into Eq.~(\ref{eq:EE1z}), we obtain
\begin{align}\label{eq:EE1H}
    \sum_{i=1}^{4} \sum_{\delta,\sigma} \sum_{\alpha,\beta=0}^{2} \tilde{\mathcal{G}}_{i,\delta,\sigma,\alpha,\beta}^{j} z^{\delta} \chi^{\sigma} \partial_{z}^{\alpha} \partial_{\chi}^{\beta} & \nonumber \\
    \times\left[\sum_{n=0}^{\mathcal{N}_z} \sum_{\mathrm{even} \, \ell}^{\mathcal{N}_{\chi}} v_{n\ell}^{i} T_{n}(z) P_{\ell}(\chi)\right] &= \sum_{\delta,\sigma} \tilde{\mathcal{S}}_{\delta,\sigma}^{j} z^{\delta} \chi^{\sigma}. 
\end{align}
By the orthogonality of Chebyshev and Legendre polynomials, we project Eq.~(\ref{eq:EE1H}) onto $T_{n'}(z)$ and $P_{\ell'}(\chi)$ to find
\begin{equation}
    \sum_{n'=0}^{\mathcal{N}_z} \sum_{\mathrm{even} \, \ell'}^{\mathcal{N}_{\chi}} w_{n'\ell'}^{j} T_{n'}(z) P_{\ell'}(\chi) = \sum_{n'=0}^{\mathcal{N}_z} \sum_{\mathrm{even} \, \ell'}^{\mathcal{N}_{\chi}} s_{n'\ell'}^{j} T_{n'}(z) P_{\ell'}(\chi), 
\end{equation}
where
\allowdisplaybreaks[4]
\begin{widetext}
\begin{align}
    w_{n'\ell'}^{j} &= \sum_{n=0}^{\mathcal{N}_z} \sum_{\mathrm{even} \, \ell}^{\mathcal{N}_{\chi}} \sum_{i=1}^{4} \sum_{\delta,\sigma} \sum_{\alpha,\beta=0}^{2} \tilde{\mathcal{G}}_{i,\delta,\sigma,\alpha,\beta}^{j} \, v_{n\ell}^{i} \int_{-1}^{+1} z^{\delta} [\partial_{z}^{\alpha} T_{n}(z)] T_{n'}(z) \frac{dz}{\sqrt{1-z^2}} \int_{-1}^{+1} \chi^{\sigma} [\partial_{\chi}^{\beta} P_{\ell}(\chi)] P_{\ell'}(\chi) \,d\chi \nonumber \label{eq:wDefinition}\\
    &\equiv \sum_{n=0}^{\mathcal{N}_z} \sum_{\mathrm{even} \, \ell}^{\mathcal{N}_{\chi}} \sum_{i=1}^{4} [\mathbb{D}_{j n' \ell', i n \ell}] v_{n\ell}^{i}, \\
    s_{n'\ell'}^{j} &= \sum_{\delta,\sigma} \tilde{\mathcal{S}}_{\delta,\sigma}^{j} \int_{-1}^{+1} z^{\delta} T_{n'}(z) \frac{dz}{\sqrt{1-z^2}} \int_{-1}^{+1} \chi^{\sigma} P_{\ell'}(\chi) \,d\chi. \label{eq:sDefinition}
\end{align}
\end{widetext}
Notice that $w_{n'\ell'}^{j}$ depend on $v_{n\ell}^i$, $M$ and $a$, while $s_{n'\ell'}^{j}$ are coefficients that depend on $c_{n\ell}$, $M$ and $a$. $\mathbb{D}_{j n' \ell', i n \ell}$ is a $4(\mathcal{N}_{z}+1)(\lfloor \mathcal{N}_{\chi}/2 \rfloor+1) \times 4(\mathcal{N}_{z}+1)(\lfloor \mathcal{N}_{\chi}/2 \rfloor+1)$ matrix, where $\lfloor x \rfloor$ is the floor function defined as the integer part of $x$.  

Since $T_{n'}$ and $P_{\ell'}$ are orthogonal functions, we must have that 
\begin{equation}\label{eq:SeparatedEq}
    w_{n'\ell'}^{j} = s_{n'\ell'}^{j}
\end{equation}
for all $n'=0,1,\ldots,\mathcal{N}_{z}$, $\ell'=0,2,\ldots,2\lfloor\mathcal{N}_{\chi}/2\rfloor$ and $j=1,2,3,4$. Equation~(\ref{eq:SeparatedEq}) is a set of linear, algebraic equations for the unknowns $v_{n\ell}^{i}$, which is effectively equivalent to the linearized, EFT-corrected field equations. Using the definition of $\mathbb{D}$ in Eq.~(\ref{eq:wDefinition}), we can write Eq.~(\ref{eq:SeparatedEq}) as a matrix equation of the form
\begin{equation}\label{eq:MatrixEq}
    \mathbb{D} \mathbf{v} = \mathbf{s}, 
\end{equation}
where the coefficient vector $\mathbf{v}$ and source vector $\mathbf{s}$ have length $4(\mathcal{N}_{z}+1)(\lfloor\mathcal{N}_{\chi}/2\rfloor+1)$ and are defined by 
\begin{align}
    \mathbf{v} &= \Big(\mathbf{v}_{0,0}^{i}, \mathbf{v}_{1,0}^{i}, \ldots, \mathbf{v}_{\mathcal{N}_z,0}^{i}, \nonumber \\
    &\qquad\qquad \ldots, \mathbf{v}_{\mathcal{N}_z,2}^{i}, \ldots, \mathbf{v}_{\mathcal{N}_{z},2\lfloor\mathcal{N}_{\chi}/2\rfloor}^{i}\Big), \\
    \mathbf{s} &= \Big(\mathbf{s}_{0,0}^{j}, \mathbf{s}_{1,0}^{j}, \ldots, \mathbf{s}_{\mathcal{N}_z,0}^{j}, \nonumber \\
    &\qquad\qquad \ldots, \mathbf{s}_{\mathcal{N}_z,2}^{j}, \ldots, \mathbf{s}_{\mathcal{N}_{z},2\lfloor\mathcal{N}_{\chi}/2\rfloor}^{j}\Big), 
\end{align}
with $i,j=1,2,3,4$ in ascending order. Thus, we have reduced a system of coupled linear partial differential equations to a linear, matrix equation. 

Recall that for the metric to asymptotically recover a BH with (ADM) mass $M$ and (ADM) angular momentum $J = M^2 a$, we need to impose the boundary conditions listed in Eq.~(\ref{eq:BoundaryCondition}). 
From Eq.~(\ref{eq:HAnsatz}), using the properties of the Chebyshev polynomials, we find that $H_i(r, \chi)$ has the asymptotic properties of 
\begin{equation}\label{eq:AsymptoticMetric}
\begin{split}
    H_i^{(0)}(\chi) &= \sum_{n=0}^{\mathcal{N}_{z}} \sum_{\mathrm{even} \, \ell}^{\mathcal{N}_{\chi}} v_{n\ell}^{i} (-1)^n P_{\ell}(\chi), \\
    H_i^{(1)}(\chi) &= 2r_+ \sum_{n=0}^{\mathcal{N}_{z}} \sum_{\mathrm{even} \, \ell}^{\mathcal{N}_{\chi}} v_{n\ell}^{i} (-1)^{n+1}n^2 P_{\ell}(\chi). 
\end{split}
\end{equation}
Note that, if one uses the compactified coordinate defined in~\cite{Stein:2014xba}, or other choices that do not involve a $1/r$ representation, the boundary conditions required would be more sophisticated than what we implement here. 
Putting Eqs.~(\ref{eq:AsymptoticMetric}) in Eq.~(\ref{eq:BoundaryCondition}), we find that 
\begin{equation}\label{eq:BoundaryConditions}
\begin{split}
    \sum_{n=0}^{\mathcal{N}_{z}} (-1)^n v_{n\ell}^{1} &= 0, \\
    \sum_{n=0}^{\mathcal{N}_{z}} (-1)^n v_{n\ell}^{2} &= r_+ \sum_{n=0}^{\mathcal{N}_{z}} (-1)^{n+1} n^2 v_{n\ell}^{3}, \\
    \sum_{n=0}^{\mathcal{N}_{z}} (-1)^n v_{n\ell}^{3} &= -2r_+ \sum_{n=0}^{\mathcal{N}_{z}} (-1)^{n+1} n^2 v_{n\ell}^{3}, \\
    \sum_{n=0}^{\mathcal{N}_{z}} (-1)^n v_{n\ell}^{4} &= -2r_+ \sum_{n=0}^{\mathcal{N}_{z}} (-1)^{n+1} n^2 v_{n\ell}^{3}, 
\end{split}
\end{equation}
for any $\ell = 0, 2, 4, \ldots, 2\lfloor \mathcal{N}_{\chi}/2 \rfloor$. 
Using these $4(\lfloor \mathcal{N}_{\chi}/2 \rfloor+1)$ constraints, we eliminate the same amount of variables, which we have chosen to remove $v_{0\ell}^{i=1,2,3,4}$ for all even $\ell$ in Eq.~(\ref{eq:MatrixEq}). 
Explicitly, we construct a reduced matrix equation 
\begin{align}\label{eq:ReducedMatrixEq}
    \tilde{\mathbb{D}} \tilde{\mathbf{v}} = \mathbf{s}, 
\end{align}
where $\tilde{\mathbb{D}}$ is a $4(\mathcal{N}_{z}+1)(\lfloor\mathcal{N}_{\chi}/2\rfloor+1) \times 4\mathcal{N}_{z}(\lfloor\mathcal{N}_{\chi}/2\rfloor+1)$ rectangular matrix constructed by 
\begin{equation}
\begin{split}
    [\tilde{\mathbb{D}}_{jn'\ell',1n\ell}] = [\mathbb{D}_{jn'\ell',1n\ell}] - (-1)^{n} [\mathbb{D}_{jn'\ell',10\ell}], \\
    [\tilde{\mathbb{D}}_{jn'\ell',2n\ell}] = [\mathbb{D}_{jn'\ell',2n\ell}] - (-1)^{n} [\mathbb{D}_{jn'\ell',20\ell}], \\
    [\tilde{\mathbb{D}}_{jn'\ell',3n\ell}] = [\mathbb{D}_{jn'\ell',3n\ell}] - (-1)^{n} [\mathbb{D}_{jn'\ell',30\ell}]\: \\
     + r_+ (-1)^{n+1} n^{2} [\mathbb{D}_{jn'\ell',20\ell}]\: \\
     - 2r_+ (-1)^{n+1} n^{2} [\mathbb{D}_{jn'\ell',30\ell}]\: \\
     - 2r_+ (-1)^{n+1} n^{2} [\mathbb{D}_{jn'\ell',40\ell}], \\
    [\tilde{\mathbb{D}}_{jn'\ell',4n\ell}] = [\mathbb{D}_{jn'\ell',4n\ell}] - (-1)^{n} [\mathbb{D}_{jn'\ell',40\ell}],
\end{split}
\end{equation}
where $n$ now runs from $1$ to $\mathcal{N}_{z}$. In addition, $\tilde{\mathbf{v}}$ now has length $4\mathcal{N}_{z}(\lfloor\mathcal{N}_{\chi}/2\rfloor+1)$, which we define as 
\begin{align}
    \mathbf{v} &= \Big(\mathbf{v}_{1,0}^{i}, \mathbf{v}_{2,0}^{i}, \ldots, \mathbf{v}_{\mathcal{N}_z,0}^{i}, \nonumber \\
    &\qquad\qquad \ldots, \mathbf{v}_{\mathcal{N}_z,2}^{i}, \ldots, \mathbf{v}_{\mathcal{N}_{z},2\lfloor\mathcal{N}_{\chi}/2\rfloor}^{i}\Big), 
\end{align}
where $i = 1,2,3,4$ is in ascending order. 
Solving this reduced matrix equation is equivalent to solving the modified field equations with boundary conditions imposed. 

In general, this rectangular system of equations might not have any solutions, given that it is an overdetermined system. Instead, we obtain the least-squares solution of the system, 
% {\ny{I'm not sure this is the right language. ``best fit''? where did you get those words from?}} \KL{changed to least-squares. I think this is the right way to say it} 
\begin{equation}
    \tilde{\mathbf{v}} = (\tilde{\mathbb{D}}^{\rm T} \tilde{\mathbb{D}})^{-1} \tilde{\mathbb{D}}^{\rm T} \mathbf{s}, 
\end{equation}
where $\tilde{\mathbb{D}}^{\rm T}$ is the transpose of $\tilde{\mathbb{D}}$. 
For all systems that we have investigated in practice, we have checked that $\mathrm{rank}(\tilde{\mathbb{D}}^{\rm T} \tilde{\mathbb{D}}) = \dim(\tilde{\mathbf{v}})$. Hence, the matrix inverse is well defined and a unique least-squares solution exists. 
To reconstruct the solution, we first use Eqs.~(\ref{eq:BoundaryConditions}) to compute $v_{0\ell}^{i=1,2,3,4}$, and then we substitute $v_{n\ell}^{i}$ into Eq.~(\ref{eq:HAnsatz}) to obtain $H_i(z,\chi)$.

\subsection{Numerical Implementation}\label{subsec:NumericalImplementation}

When constructing the scalar field solution numerically, for simplicity, we choose to truncate the mode expansion at some spectral order $N_z = N_{\chi} = N$.  
Since the metric corrections are sourced from the scalar field, we would need to insert the scalar field solution, through the coefficients $c_{n\ell}$, back into the EFT-corrected field equations to obtain $H_i$. 
In our numerical implementation, we will choose to insert the scalar field solution computed at $N = 50$ into Eq.~(\ref{eq:ReducedMatrixEq}). 
For all spins up to $a \leq 0.999$, the absolute error of the scalar field equations (defined in Eq.~(\ref{eq:ScalarFieldResidual})) truncated at $N = 50$ is $\lesssim 10^{-12}$, which is sufficiently accurate to calculate the metric corrections to the precision sought in this paper. 
Similarly, we also set $\mathcal{N}_z = \mathcal{N}_{\chi} = \mathcal{N}$ to simplify our code implementation. 

To retain numerical precision, we compute $\mathfrak{D}_{\ell}$ and $\mathfrak{s}_{\ell}$ with a numerical precision of 256 digits.
To invert $\mathfrak{D}_{\ell}$ and $\tilde{\mathbb{D}}^{\rm T} \tilde{\mathbb{D}}$, we use the built-in function \texttt{Inverse} in \emph{Mathematica} with the same numerical precision. 
We have checked that further increasing the numerical precision does not modify the solution and other physical quantities computed in the rest of the paper to the precision sought in this paper (as described in the next section). 

%----------------------------------------------------------------
\section{Background scalar field : results}\label{sec:ScalarFieldResult}

\begin{figure*}[t!]
    \centering
    \includegraphics[width=\textwidth]{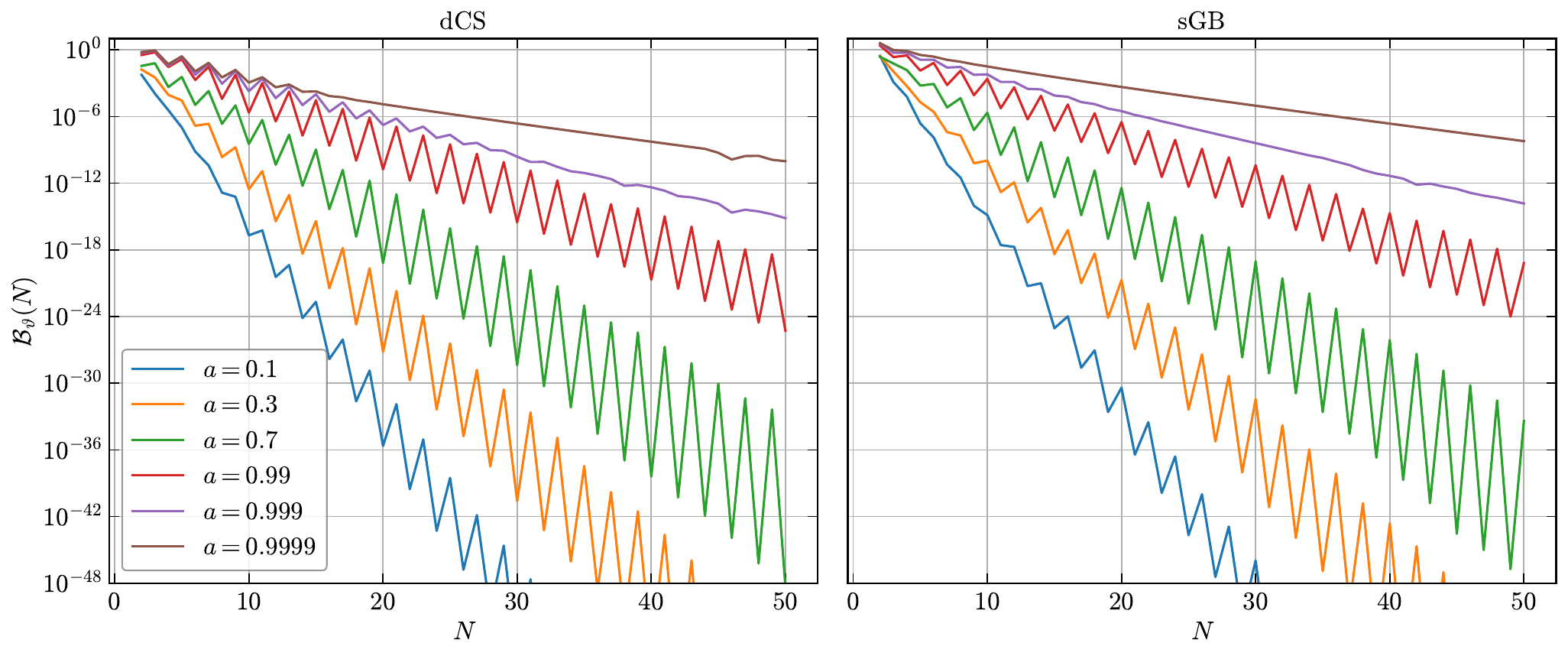}
    \caption{Scalar field backward modulus difference $\mathcal{B}_{\vartheta}(N)$, defined by Eq.~(\ref{eq:BWD_scalar_field}) in the main text, of the sGB and dCS scalar field solution as a function of spectral order $N$ computed at different values of spin $a$. }
    \label{fig:ScalarFieldBMD}
\end{figure*}

First, we examine the convergence behavior of the scalar field solution with $N$. Since we are using the Chebyshev basis functions with ``roots'' grid points, we expect exponential convergence \cite{boyd2013chebyshev}. 
To quantitatively describe the convergence behavior, we define the scalar field backward modulus difference $\mathcal{B}_{\vartheta}(N)$ as 
\begin{equation}\label{eq:BWD_scalar_field}
    \mathcal{B}_{\vartheta}(N) = \left[ \int_{r_+}^{\infty} \int_{-1}^{+1} \frac{(\vartheta^{N} - \vartheta^{N-1})^2}{r^2}\sqrt{-g^{(0)}} \, dr d\chi \right]^{1/2}, 
\end{equation}
where $\vartheta^{N}(r,\chi)$ represents the scalar field solution computed at spectral order $N$, and $\sqrt{-g^{(0)}} = \Sigma$ is the determinant of the Kerr metric. 
This quantity estimates the $L^2$ norm of the scalar field residual weighted by $1/r^2$. The extra weight of $1/r^2$ is convenient to regularize the integral at spatial infinity, although any weight of the form $1/r^p$ with $p>1$ would work. What matters here is the behavior of ${\cal{B}}_{\theta}$ with the spectral order $N$, as it can reveal exponential spectral convergence for different values of spin.  

We plot the backward modulus difference as a function of spectral order $N$ in Fig. \ref{fig:ScalarFieldBMD}. 
Individual lines represent the convergence behavior of solutions with different spins. 
Observe that $\mathcal{B}_{\vartheta}(N)$, in general, decreases at an exponential rate, which validates our expectation. 
Observe also that there is an oscillation pattern in Fig.~\ref{fig:ScalarFieldBMD} when $a \leq 0.99$. 
For example, in dCS gravity, $\mathcal{B}_{\vartheta}(N)$ for even $N$ is smaller than that at the adjacent odd spectral orders.
This suggests the additional terms from the $N$-th spectral order, i.e., terms proportional to $T_{N}(z)P_{\ell}(\chi)$ and $T_{n}(z)P_{N}(\chi)$ ($n,\ell \leq N$), are small compared to those from the $(N-1)$-th spectral order. 
Since $N$ is even, terms with $P_{N}(\chi)$ vanish due to our ansatz, leaving only contributions from $T_{N}(z)P_{\ell}(\chi)$. 
Hence, $\mathcal{B}_{\vartheta}(N)$ is small compared to $\mathcal{B}_{\vartheta}(N-1)$. In sGB gravity, we observe a similar pattern, where we find $\mathcal{B}_{\vartheta}(N)$ at odd $N$ is smaller than at even $N$. Once again, this can be explained by the parity structure of the ansatz that we adopted. 

When $a \geq 0.999$, the oscillation patterns of Fig.~\ref{fig:ScalarFieldBMD} diminish. 
This is because the contribution from the $T_{N}(z)P_{\ell}(\chi)$ terms now becomes significant.
This can be understood from Eq.~(\ref{eq:ScalarFieldSource}), where the source term involves hypergeometric functions, which can be represented as an infinite series in $a^2/r^2$.
We expect the scalar field solution can also be represented as an infinite series in $a^2/r^2$. 
When $a$ is small, the $1/r^{N}$ terms are suppressed relative to lower order terms, such as $1/r^{N-1}$, due to the extra factor of $a$. 
As $a$ increases, the $1/r^{N}$ terms become comparable to the other lower-order terms. 
Projecting these $1/r^{N}$ terms onto Chebyshev polynomials, they become terms proportional to $T_N(z)$.  
Hence, the $T_{N}(z)P_{\ell}(\chi)$ terms are no longer small. 
and $\mathcal{B}_{\vartheta}(N)$ decreases smoothly as $N$ increases, matching what we see in sequences with $a \geq 0.999$.

Now, let us examine slowly-rotating BHs ($a \lesssim 0.1$). 
We see that $\mathcal{B}_{\vartheta}(N)$ can easily go below machine precision even at small spectral order $N \approx 10$. 
Since series-in-$a$ solutions in dCS and sGB gravity can be written as a series in $T_{n}(z)$ and $P_{\ell}(\chi)$ exactly, with a finite and low number of terms \cite{Cano_Ruiperez_2019, Yagi:2012ya}, we only need a small number of spectral bases to capture most of the features of the solution. 
As we increase the spin, higher-order multipole moments become more and more manifest; thus, a higher spectral order is required to acquire the same precision.
This result matches that shown in~\cite{Stein:2014wza}, despite our choice of a different compactified coordinate. 
Near the extremal limit, at $a = 0.9999$, although the convergence rate is smaller than at smaller spins, we find that $\mathcal{B}_{\vartheta}(N = 50) \sim 10^{-9}$ for both sGB and dCS gravity, indicating the scalar field solutions are sufficiently precise for our further computations. 

\begin{figure}[t!]
    \centering
    \includegraphics[width=\columnwidth]{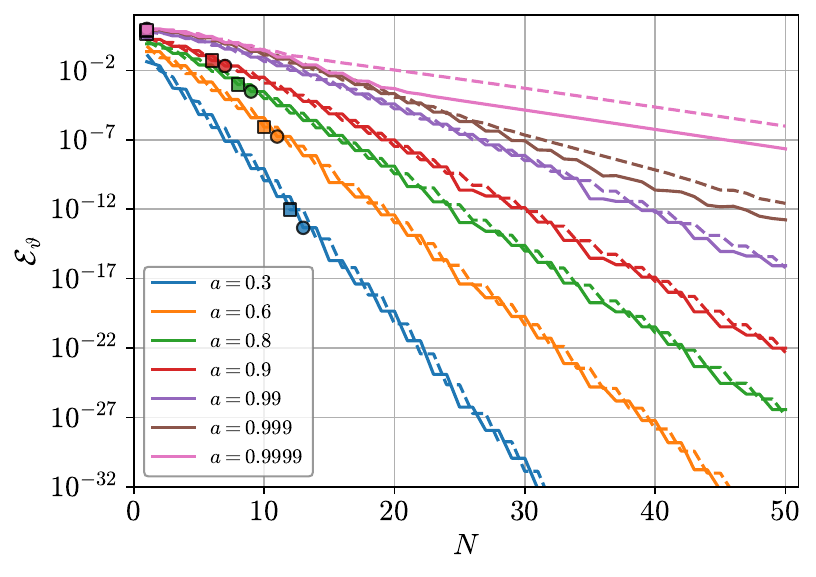}
    \caption{The absolute error of the sGB and dCS scalar field equation against spectral order $N$. 
    Each solid (dashed) line represents the trends of dCS (sGB) scalar field at different values of spin. 
    The circles (squares) on each line mark the spectral order at which the residual error of the spectral dCS (sGB) scalar field becomes lower than that of the series-in-$a$ solution up to ${\cal{O}}(a^{15})$. }
    \label{fig:ScalarFieldResidual}
\end{figure}

Next, to study the precision to which the spectral solution satisfies Eq.~(\ref{eq:SF0}), we define the absolute error $\mathcal{E}_{\vartheta}$ by
\begin{equation}\label{eq:ScalarFieldResidual}
    \mathcal{E}_{\vartheta} = \left[ \int_{r_+}^{\infty} \int_{-1}^{+1} E_{\vartheta}^{2} \sqrt{-g^{(0)}} \, dr d\chi \right]^{1/2}, 
\end{equation}
which is the $L^2$ norm of Eq.~(\ref{eq:SF0}), with $\sqrt{-g^{(0)}}$ included as the Jacobian of the GR metric. 
The absolute error is meant to characterize the accuracy of the spectral solution over the whole spacetime\footnote{Although the absolute error is an integral over space, since the background scalar field that we construct is stationary, the integral also characterizes the error over the whole spacetime.}. Even though we do not need to multiply $E_{\theta}^2$ by a regularization factor, what will matter to us is the behavior of this $L^2$ norm as a function of spectral order $N$ and spin.

In Fig.~\ref{fig:ScalarFieldResidual}, we show $\mathcal{E}_{\vartheta}$ as a function of spectral order $N$. 
First, as we increase the spectral order, $\mathcal{E}_{\vartheta}$ decreases approximately exponentially for both sGB and dCS gravity, a result that is consistent with the behavior observed in $\mathcal{B}_{\vartheta}(N)$.
Instead of having the rapidly oscillating pattern in Fig.~\ref{fig:ScalarFieldBMD}, we find that $\mathcal{E}_{\vartheta}$ decreases steadily. 
Specifically, for the slowly-rotating BHs in dCS gravity, we find that the even $N$-th spectral order solution always has a similar $\mathcal{E}_{\vartheta}$ as the odd $(N-1)$-th spectral order solution. 
This is consistent with our explanation in the previous paragraphs, where the additional $T_N(z)$ term in $\tilde{\vartheta}_{\ell}$ does not modify the solution greatly when $a \leq 0.99$. 
For sGB solutions, the opposite occurs where the odd $N$-th spectral order solution has a similar $\mathcal{E}_{\vartheta}$ as the even $(N-1)$-th spectral order solution. 
For solutions with $a \geq 0.999$, following the same argument as before, the pattern disappears because the $T_{N}(z)$ term becomes large, and $\mathcal{E}_{\vartheta}$ decreases monotonically as $N$ increases. 
We limit the maximum spin to $a = 0.9999$, where the absolute error at $N = 50$ is around $10^{-6}$.
If one were to be interested in even more rapidly-rotating BHs, one would either have to keep more terms in the expansions, or find a more efficient representation of the solution.

One interesting behavior shown in Fig.~\ref{fig:ScalarFieldBMD} and \ref{fig:ScalarFieldResidual} is that the dCS $\mathcal{B}_{\vartheta}(N)$ and $\mathcal{E}_{\vartheta}$ approximately overlap with the sGB ones. 
This arises from the fact that Eq.~(\ref{eq:SF0}) for sGB and dCS gravity are \emph{dual} to each other. 
This can be seen from the fact that $s_{\ell, q=1}(r)$ and $s_{\ell, q=2}(r)$ are the real and imaginary parts of the same complex function. 
If we were to consider another beyond-Einstein gravity theory, the convergence behavior and absolute error could be different from what we have here. 

\begin{figure}[t!]
    \centering
    \includegraphics[width=\columnwidth]{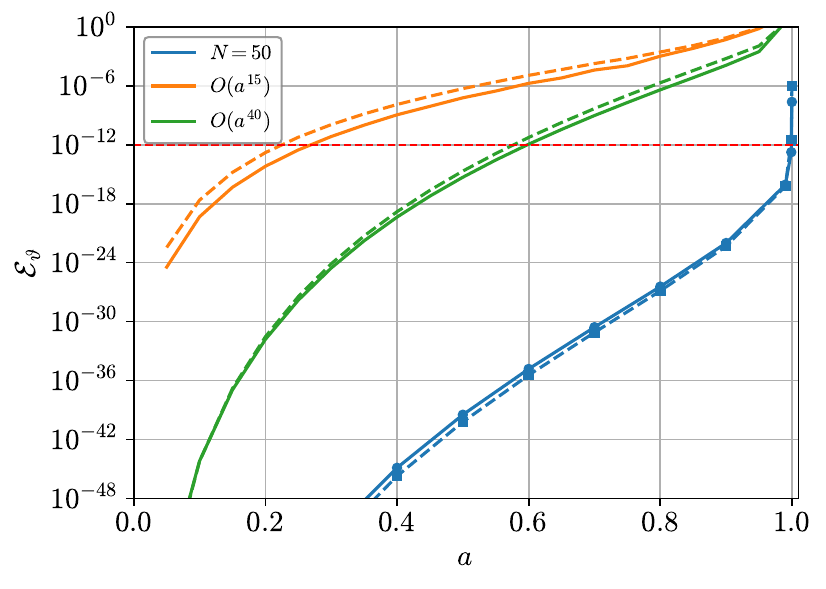}
    \caption{The absolute error of the sGB (dashed lines) and dCS (solid lines) scalar field equation computed for the spectral and series-in-$a$ solution. The $N = 50$ spectral solutions (blue) are taken to compare with the series-in-$a$ solutions up to ${\cal{O}}(a^{15})$ (orange) and ${\cal{O}}(a^{40})$ (green). }
    \label{fig:ScalarFieldResidual50}
\end{figure}

Next, we compare the accuracy of the spectral solution with the series-in-$a$ solution found in~\cite{Cano_Ruiperez_2019}. 
Since the latter solution is computed by assuming \textit{a priori} a slow-rotation expansion (in the form of a Taylor series in $a \ll 1$), it is expected that, when the spin is large, that solution will cease to be accurate. Figure~\ref{fig:ScalarFieldResidual50} shows the absolute error of the dCS (sGB) series-in-$a$ solution up to ${\cal{O}}(a^{15})$ (calculated in~\cite{Cano_Ruiperez_2019}) and the extended series-in-$a$ solution up to ${\cal{O}}(a^{40})$ (calculated in~\cite{Chung:2024ira, Chung:2024vaf, Chung:2025gyg}), which are to be compared to the spectral solutions computed in this paper with $N = 50$.
Observe that the absolute error of the spectral solution is always below the series-in-$a$ solutions by $\gtrsim 10$ orders of magnitude.
The series-in-$a$ solution up to ${\cal{O}}(a^{15})$ generally has a larger error than the one computed to ${\cal{O}}(a^{40})$, except in cases near $a = 1$, where both solutions are inaccurate. 
As spin increases, to achieve the same accuracy as the series-in-$a$ solutions, a higher spectral order is required. 
For near-extremal solutions, we find that the absolute error of both the spectral and series-in-$a$ solutions increases rapidly, though the $\mathcal{E}_{\vartheta}$ of the spectral solution is still relatively small at around $10^{-6}$. 
This ``almost'' divergent behavior suggests that the ansatz we chose is not the best one to represent extremal solutions, but highly accurate for near extremal BHs and BHs with smaller spins. 

Recall that AD gravity has two fields, a scalar one that behaves exactly the same as the sGB scalar, and a pseudoscalar one that behaves exactly the same as the dCS scalar. 
Therefore, the absolute error of the AD field equations would then be the sum of that of the sGB and the dCS solution, and thus, of the same order of magnitude and with the same spin dependence. 
AD gravity will differ from sGB and dCS gravity in how these scalars back-react on the metric, especially when considering non-linear coupling interactions, but such will be higher-order in the small-coupling approximation, as we shall see in the next section. 

Figure~\ref{fig:ScalarFieldResidual50} clearly indicates that, for a fixed spin, there is some $N$ at which the spectral solution becomes as accurate as the series-in-$a$ solution; for any $N$ larger than this, the spectral solution would become more accurate. This is shown in Fig.~\ref{fig:ScalarFieldResidual} through circle (square) markers.
Observe that, for small $a$, the series-in-$a$ solution is as accurate as the spectral solution computed at $N = 13$. 
As we continue to increase $a$, the accuracy of the series-in-$a$ solution decreases rapidly. 
Even when keeping terms up to ${\cal{O}}(a^{40})$, the series-in-$a$ is only valid up to $a \approx 0.6$ before the spectral solution becomes more accurate.
This observation reveals the limitation of slow-rotation expansion in constructing accurate space-time of nearly extremal black holes.

Suppose now that we wished to obtain solutions to a given, fixed accuracy, for example, $\mathcal{E}_{\vartheta} = 10^{-12}$. As shown in Fig.~\ref{fig:ScalarFieldResidual50}, the series-in-$a$ solution up to ${\cal{O}}(a^{15})$ would then only be valid for spins $a \lesssim 0.3$. 
Considering that observations of binary BH merger remnants or supermassive BHs have spins typically larger than $0.6$ \cite{KAGRA:2021vkt, Reynolds:2013rva}, if one uses the series-in-$a$ solution in data analysis, systematic error would largely bias the result. 
Meanwhile, the spectral solution is valid up to $a \approx 0.99$, which includes most of the observations to date. 
This means the spectral solutions presented here can be used in many current and future observations without introducing as much systematic bias to data analysis tasks. Our discussion here has focused on the scalar field only (which is not directly observable), but the same conclusions hold for the metric itself, as we show in the next section. 

The number of terms in the scalar field expression directly impacts the computational costs and efficiency, so let us compare the length of the spectral solution to that of the series-in-$a$ solution. 
As an example, at $a = 0.6$, the series-in-$a$ solution up to ${\cal{O}}(a^{40})$ contains 250 terms in total. 
On the other hand, the spectral solution with the same $\mathcal{E}_{\vartheta}$ has 200 terms in total, a 20\% decrease in length. 
As $a$ increases, the difference in length becomes larger, as a longer series-in-$a$ solution is required to maintain similar accuracy as the spectral solution. 
Thus, the computational cost of tasks that utilize spectral solutions will be lower than those that employ series-in-$a$ solutions, especially for larger in spin cases.

\begin{figure*}[t!]
    \centering
    \subfloat{\includegraphics[width=0.9\textwidth]{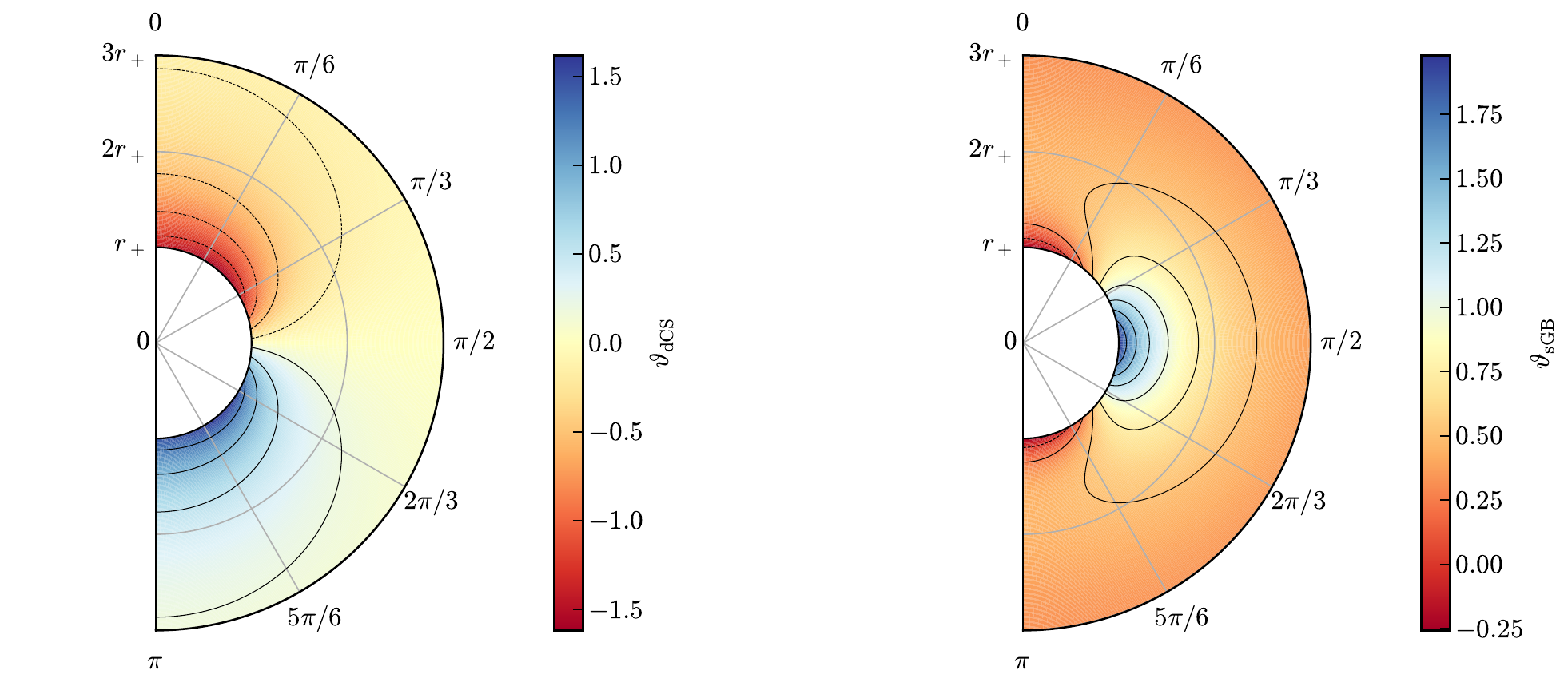}}
    \caption{Meridional cross-section of the scalar field profiles for sGB (right) and dCS gravity (left) at $a = 0.9$, computed at spectral order $N = 50$. The color bars indicate the value of the scalar fields, while the contour lines illustrate their multipolar structure. Observe that the dCS scalar field is odd in parity, while the sGB scalar field is even in parity, as expected.}
    \label{fig:dCS_sGB_ScalarField}
\end{figure*}

We conclude this section by showing a rapidly rotating scalar field profile in sGB and dCS gravity. 
Figure~\ref{fig:dCS_sGB_ScalarField} shows the cross-section of $\vartheta_{\rm dCS/sGB}$ in the $r-\chi$ plane computed with $N = 50$. 
Observe that $\vartheta_{\rm dCS}$ has a dominant dipole structure, matching that found in \cite{Stein:2014wza}\footnote{Our results match those of~\cite{Stein:2014xba} in the dCS case, but, due to a different definition of the coupling constant, the dCS 
    scalar fields differ by a factor of $-1/8$.}. 
In the sGB case, other than the leading monopole charge that we see, we also find a quadrupole pointing along the polar axis. 
Such higher multipole structure becomes larger and larger as $a$ increases, leading to substantial changes in the scalar field profiles of rapidly-rotating black holes.

%----------------------------------------------------------------
\section{Metric Corrections: Results}\label{sec:MetricCorrectionsResult}

In this section, we examine the metric corrections of sGB and dCS gravity. 
Similar to the case of scalar fields, we assess the convergence behavior and absolute error of the metric corrections first. 
Then, we compare our spectral solutions to existing series-in-$a$ solutions for validation, and extend the results to rapidly rotating cases. 

\begin{figure*}[t!]
    \centering
    \includegraphics[width=\textwidth]{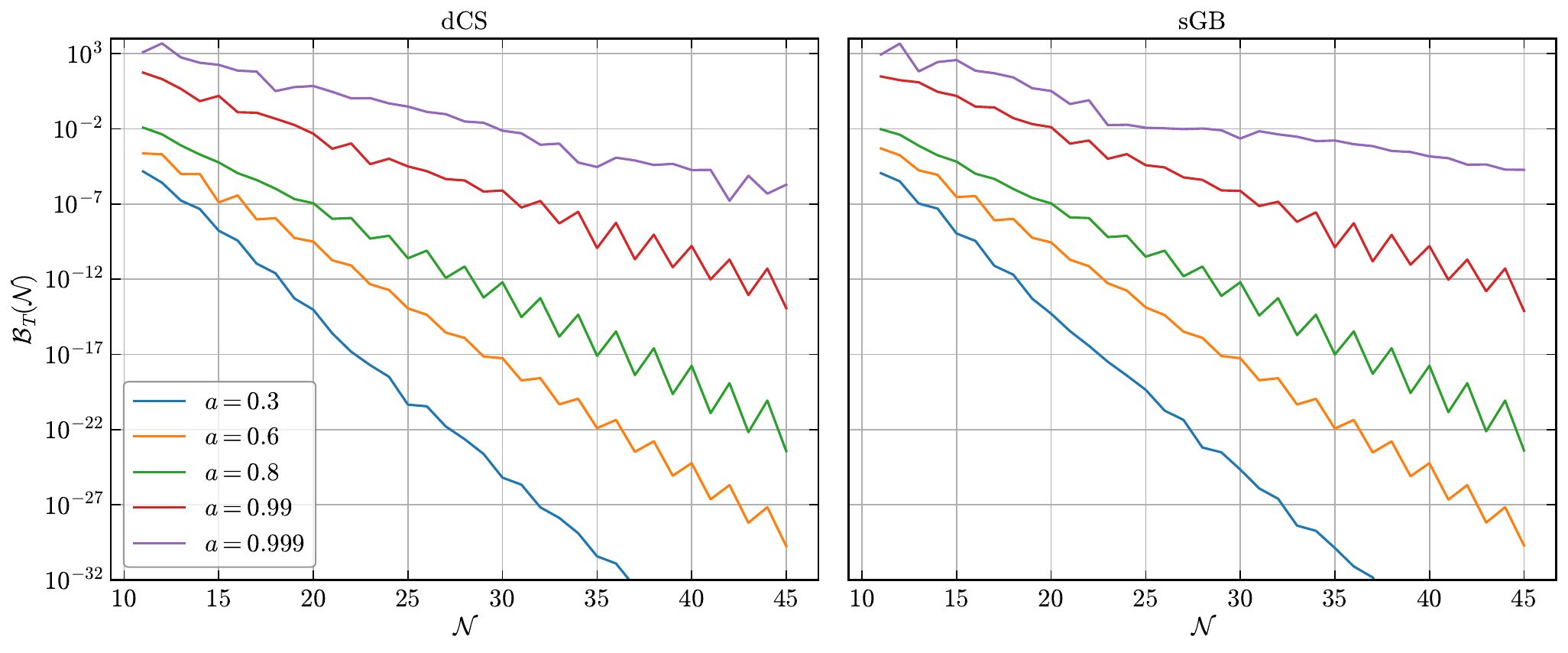}
    \caption{The total backward modulus difference of the metric corrections $\mathcal{B}_T(\mathcal{N})$ in dCS (left) and sGB (right) gravity as a function of spectral order $\mathcal{N}$ computed at different values of spin $a$. }
    \label{fig:EEBMD}
 \end{figure*}

First, we wish to verify exponential convergence. 
To this end, we define the backward modulus difference for metric corrections, 
\begin{align}
    \mathcal{B}_{i}(\mathcal{N}) = \left[ \int_{r_+}^{\infty} \int_{-1}^{+1} \frac{(H_i^{\mathcal{N}} - H_i^{\mathcal{N}-1})^2}{r^4} \sqrt{-g^{(0)}}\,drd\chi \right]^{1/2}, 
\end{align}
and the total backward modulus difference for metric corrections, 
\begin{align}
    \mathcal{B}_{T}(\mathcal{N}) = \sum_{i=1}^{4}  \mathcal{B}_{i}(\mathcal{N})  , 
\end{align}
where $H_i^{\mathcal{N}}(r,\chi)$ represents $H_i(r, \chi)$ computed at spectral order $\mathcal{N}$. 
The backward modulus $\mathcal{B}_i(\mathcal{N})$ is the $L^2$ norm of metric correction residuals weighted by $1/r^4$. As in the case for the $L^2$ norm of the scalar field, the $1/r^4$ factor is introduced so that the $L^2$ norms ${\cal{B}}_i$ converge at spatial infinity. As such, what matters is its behavior with the spectral order ${\cal{N}}$ and its relative size for different choices of spin. 

We plot $\mathcal{B}_T(\mathcal{N})$ against spectral order $\mathcal{N}$ in Fig.~\ref{fig:EEBMD} for a set of spins up to $a=0.999$. 
Note that $\mathcal{B}_{T}(\mathcal{N})$ converges if and only if all $\mathcal{B}_{i}(\mathcal{N})$ converge. 
As in the scalar field case, individual sequences representing different spins exhibit approximate exponential convergence. 
In particular, spectral solutions with small $a$ have a faster rate of convergence, and the convergence rate decreases as spin increases.
We also attempted to compute solutions for $a > 0.999$, however, we found the convergence rate unsatisfactory, which is why we here restrict attention to $a \leq 0.999$.

\begin{figure}[t!]
    \centering
    \includegraphics[width=\columnwidth]{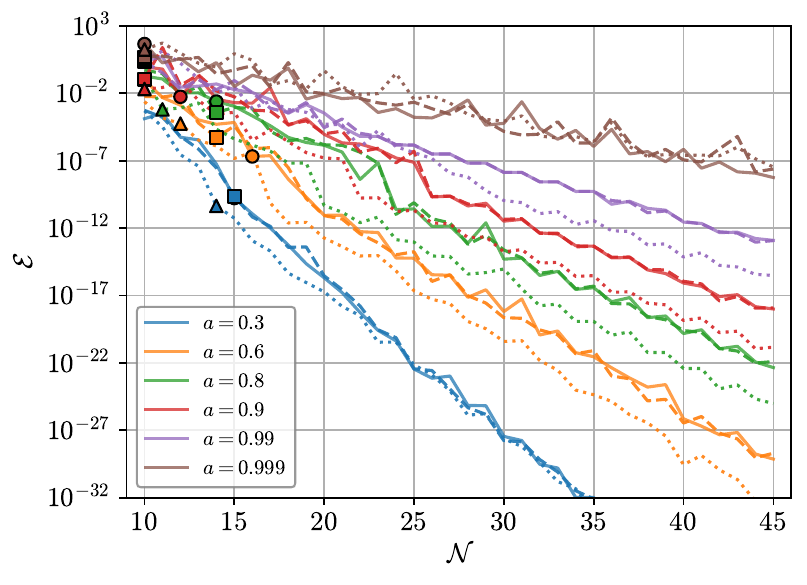}
    \caption{Absolute error of the EFT-corrected field equations against spectral order $\mathcal{N}$. The solid, dashed and dotted lines represent solutions with different values of spin in dCS, sGB and AD gravity, respectively. The circles, squares and triangle on each line mark the spectral order at which the absolute error computed with our dCS, sGB and AD solution becomes lower than that calculated with the series-in-$a$ solution with terms up to ${\cal{O}}(a^{15})$}
    \label{fig:EEResidual}
\end{figure}

Let us now assess the accuracy of the metric corrections. No unique way exists to estimate the residual of an approximate solution of a given equation to our knowledge, so we here define the absolute error of the field equations as
\begin{equation}\label{eq:EEResidual}
    \mathcal{E} = \left[ \int_{r_+}^{\infty} \int_{-1}^{+1} E_{\mu\nu}E^{\mu\nu} \frac{\Delta^4}{r^8} (1 - \chi^2)^2 \sqrt{-g^{(0)}}\,drd\chi \right]^{1/2}.
\end{equation}
We regularize the integral by a factor of $\Delta^4 r^{-8} (1-\chi^2)^2$ in order to prevent divergences at the horizon and the poles. 
This factor is a simple rational function constructed such that $\mathcal{E}$ converges, and thus, what matters is its relative size with different choices of the spectral order ${\cal{N}}$ and spin. 

Figure~\ref{fig:EEResidual} shows the absolute error of the field equations $\mathcal{E}$ against spectral order $\mathcal{N}$. 
For $\mathcal{N} < 10$, we expect that the spectral solution will not be as good as an approximation as the series-in-$a$ solution, since the latter, at leading order in $a$, contains terms up to ${\cal{O}}(r^{-9})$; 
hence, we start our calculations at spectral order $\mathcal{N} = 10$. 
Overall, we find that the absolute error decreases approximately exponentially for all quadratic gravity solutions, as in the case of the scalar fields. 
Observe that spectral solutions with small spin converge faster than the ones with large spin.
This is consistent with the backward modulus difference we studied in Fig.~\ref{fig:EEBMD}. 
For all BHs with $a\leq 0.999$, the spectral solutions with $\mathcal{N} = 45$ have an absolute error $\lesssim 10^{-8}$, which ensures that all solutions we obtain are sufficiently accurate for our purposes. One can easily obtain more accurate solutions by choosing a larger value of $\mathcal{N}$ (and $N$).

Let us now compare the spectral solutions to the series-in-$a$ solution of~\cite{Cano_Ruiperez_2019}.
Figure~\ref{fig:EEResidual} contains markers on each sequence that indicate the minimal spectral order at which the series-in-$a$ solution computed up to ${\cal{O}}(a^{15})$ has a larger $\mathcal{E}$ than the spectral solution. 
Specifically, when $a = 0.3$, the absolute error of the series-in-$a$ solution is comparable to that of the $\mathcal{N} = 15$ spectral solution. 
For larger $a$, a spectral solution with smaller $\mathcal{N}$ could attain the same accuracy as the series-in-$a$ solution. 
For near-extremal solutions, the markers all lie around $\mathcal{N} = 10$, with an error exceeding ${\cal{O}}(1)$. 
This indicates the series-in-$a$ solution does not satisfy Eq.~(\ref{eq:EE1}) at all, and that the slow-rotation Taylor expansion simply fails in this regime. 

\begin{figure}[t!]
    \centering
    \includegraphics[width=\columnwidth]{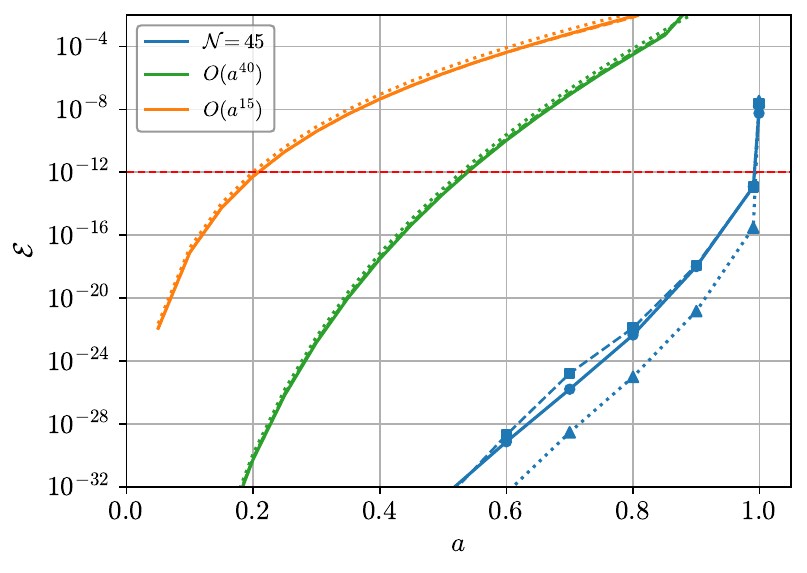}
    \caption{The absolute error of the modified Einstein field equations in the dCS (solid lines), sGB (dashed lines) and AD (dotted lines) gravity, computed for the spectral and series-in-$a$ solution. 
    The $\mathcal{N} = 45$ spectral solutions are taken to compare with the series-in-$a$ solutions computed up to ${\cal{O}}(a^{15})$ and ${\cal{O}}(a^{40})$.}
    \label{fig:EEResidual45}
\end{figure}

\begin{figure*}[t!]
    \centering
    \subfloat{\includegraphics[height=0.225\textheight]{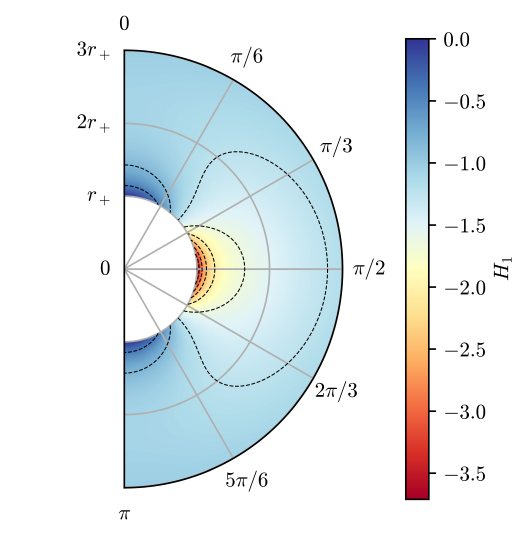}}
    \hspace{1cm} %
    \subfloat{\includegraphics[height=0.225\textheight]{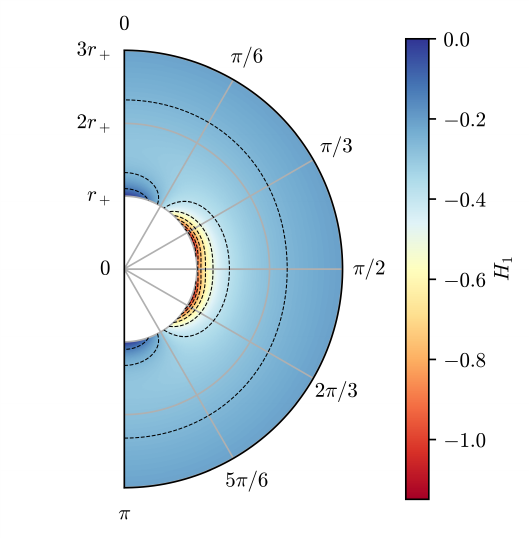}}
    \hspace{1cm} %
    \subfloat{\includegraphics[height=0.225\textheight]{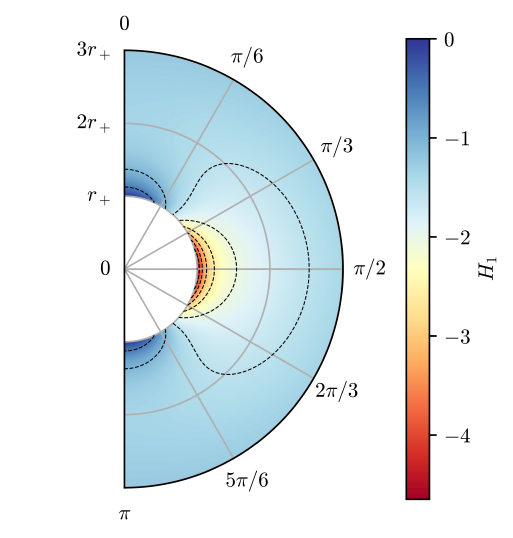}}
    \\
    \subfloat{\includegraphics[height=0.225\textheight]{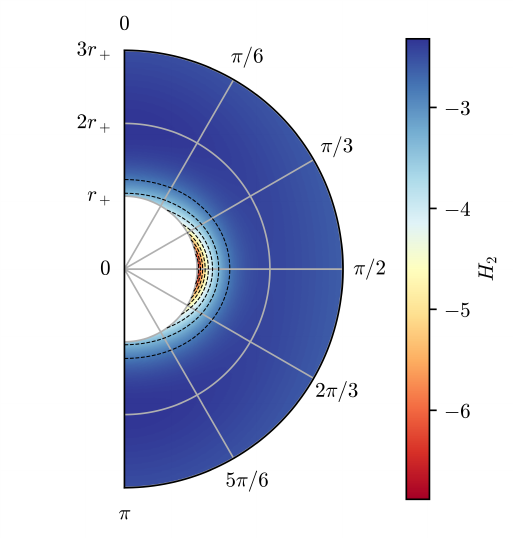}}
    \hspace{1cm} %
    \subfloat{\includegraphics[height=0.225\textheight]{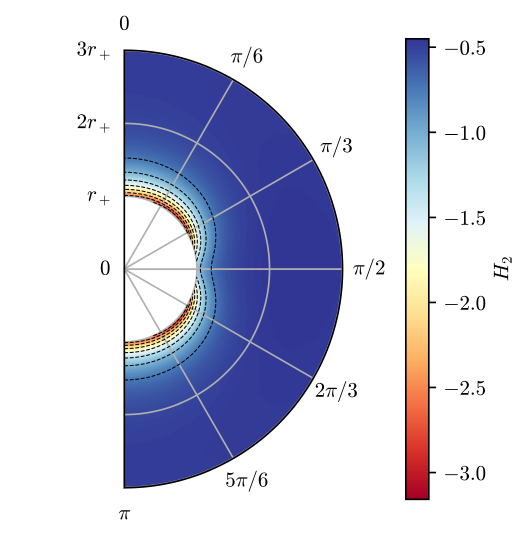}}
    \hspace{1cm} %
    \subfloat{\includegraphics[height=0.225\textheight]{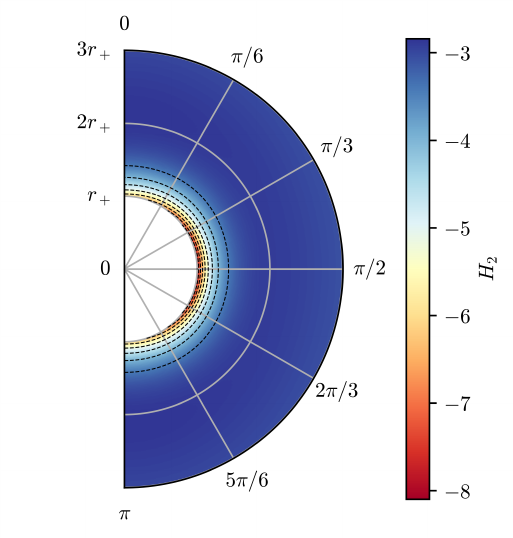}}
    \\
    \subfloat{\includegraphics[height=0.225\textheight]{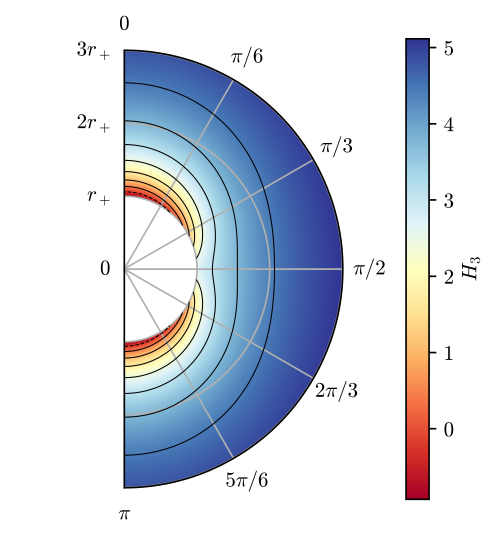}}
    \hspace{1cm} %
    \subfloat{\includegraphics[height=0.225\textheight]{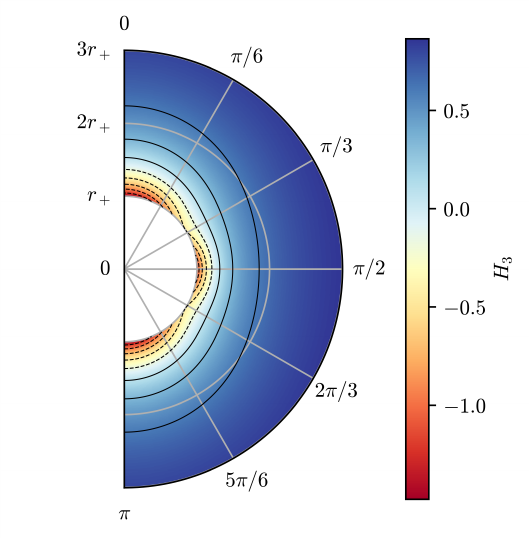}}
    \hspace{1cm} %
    \subfloat{\includegraphics[height=0.225\textheight]{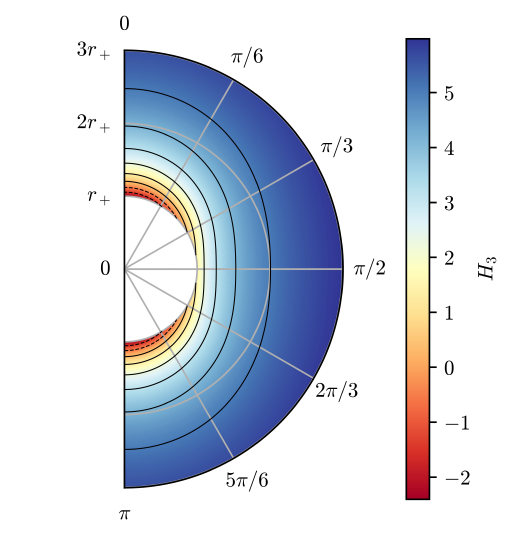}}
    \\
    \subfloat{\includegraphics[height=0.225\textheight]{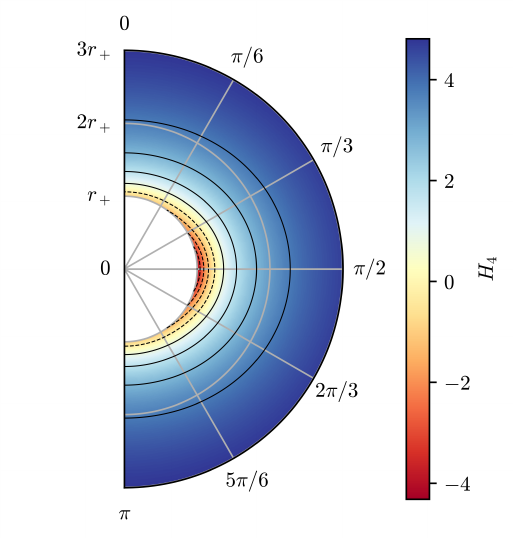}}
    \hspace{1cm} %
    \subfloat{\includegraphics[height=0.225\textheight]{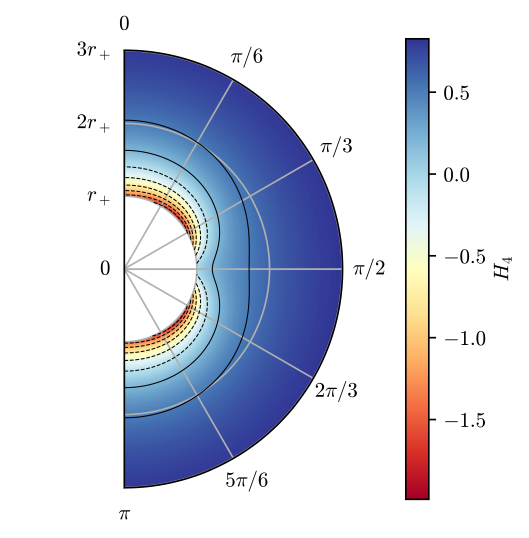}}
    \hspace{1cm} %
    \subfloat{\includegraphics[height=0.225\textheight]{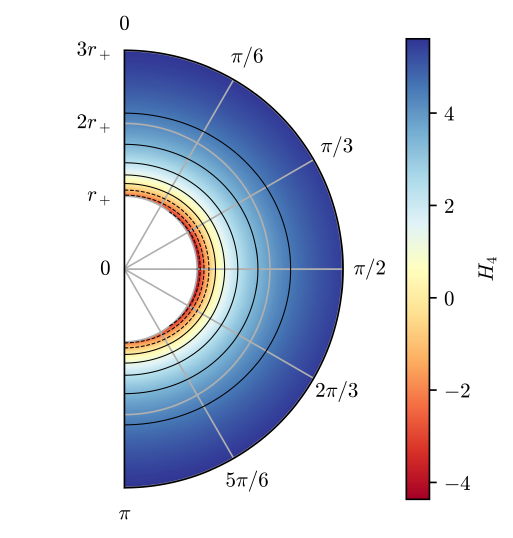}}
    \caption{The metric corrections of a rotating black hole with spin $a = 0.9$ in dCS (left), sGB (middle) and AD (right) gravity, computed at spectral order $\mathcal{N} = 45$. 
    Contour lines highlight the multipolar structure of $H_i(r, \chi)$. }
    \label{fig:dCS_sGB_Metric_Corrections}
\end{figure*}

We now compare our spectral solutions (at $\mathcal{N} = 45$) to the series-in-$a$ solutions to determine the regime of validity of the slow-rotation expansion in more detail. 
Figure~\ref{fig:EEResidual45} shows the absolute error of the spectral solutions and the series-in-$a$ solutions to ${\cal{O}}(a^{15})$ and ${\cal{O}}(a^{40})$ against spin. 
In general, both of the series-in-$a$ solutions have an error much larger than the spectral solution. 
The series-in-$a$ solution to ${\cal{O}}(a^{15})$ generally has $\mathcal{E}$ larger than that of the series-in-$a$ solution to ${\cal{O}}(a^{40})$, except for rapidly rotating solutions, where both solutions fail to approximate the true metric corrections. 
Taking the same tolerance as that chosen in Sec.~\ref{sec:ScalarFieldResult}, $\mathcal{E}_{\rm max} = 10^{-12}$, we find that the series-in-$a$ solution computed up to ${\cal{O}}(a^{15})$ (${\cal{O}}(a^{40})$) is only accurate up to $a \approx 0.2$ ($a \approx 0.5$). 
Meanwhile, the spectral solution can withstand such a tolerance up to $a \approx 0.99$.
Even when we extend the spectral solution to the Thorne limit of $a \approx 0.998$ \cite{Thorne:1974ve}, the spectral solution maintains an error of ${\cal{O}}(10^{-8})$. 
This implies that the spectral solution is an excellent approximation to the true solution, and that it can be used to compare observables computed from it to all astrophysical BH observational data, including rapidly rotating supermassive BHs \cite{Reynolds:2013rva}, without incurring mismodeling bias. 

To assess the computational efficiency of tasks that utilize $H_i$, we compare the length of the spectral solution and the series-in-$a$ solution to ${\cal{O}}(a^{40})$. 
In the series-in-$a$ solution, at $a = 0.6$, the metric components contain terms up to $1/r^{46}$ and $\chi^{40}$, with a total of 357 terms. 
Meanwhile, the spectral solution with the same accuracy (achieved with $\mathcal{N} = 19$) only has terms up to $1/r^{19}$ and $\chi^{19}$, containing 200 terms in total. 
Hence, for both the scalar field solution and the metric corrections, it is more computationally efficient to use the spectral solution instead of the series-in-$a$ solution. 

Lastly, we examine the multipolar structure of the sGB and dCS metric corrections. 
We present, for the first time, the metric corrections of rapidly rotating BH solutions in sGB and dCS gravity in  Fig.~\ref{fig:dCS_sGB_Metric_Corrections} for $a = 0.9$. 
We show the meridional cross-sections of the metric corrections in Fig.~\ref{fig:dCS_sGB_Metric_Corrections}, where sGB, dCS and AD corrections are shown. 
Note that the metric corrections are even in $\chi$, as implied by the parity symmetry of both gravity theories. 
For $H_1$ (top panels), we can see a clear quadrupole moment along the polar axis. 
Since $H_1$ is the correction to $g_{tt}$, the ergosphere, defined as the surface where $g_{tt} = 0$, would be altered considerably by a quadrupole contribution. 
Other metric corrections also modify the spacetime exterior to the BH, leading to corrections of physical observables.

%----------------------------------------------------------------
\section{Closed-form analytic black hole solution} \label{sec:FittedSolution}

Having BH solutions at a wide range of spins, we now construct a fully closed-form/analytic solution that satisfies the field equations to leading-order in the EFT coupling to extremely high precision, by fitting the individual spectral coefficients as functions of the spin $a$. 
Unlike interpolation, our fitting function takes into account the global behavior of the solutions. 
We first devise a general fitting ansatz that could capture the spin dependence of the expansion coefficients. 
Then, we construct fitted scalar fields and metric corrections using spectral solutions at $N = 50$ and $\mathcal{N} = 45$, and their absolute errors are subsequently evaluated. 
This solution, to our knowledge, is the first closed-form, analytic, rapidly rotating BH solution in beyond-Einstein gravity. 
The analytic solutions of the scalar fields and metric corrections are provided in the supplementary material as a \emph{Mathematica} notebook.

\subsection{Scalar Fields}

\begin{figure*}[t!]
    \centering
    \includegraphics[width=\textwidth]{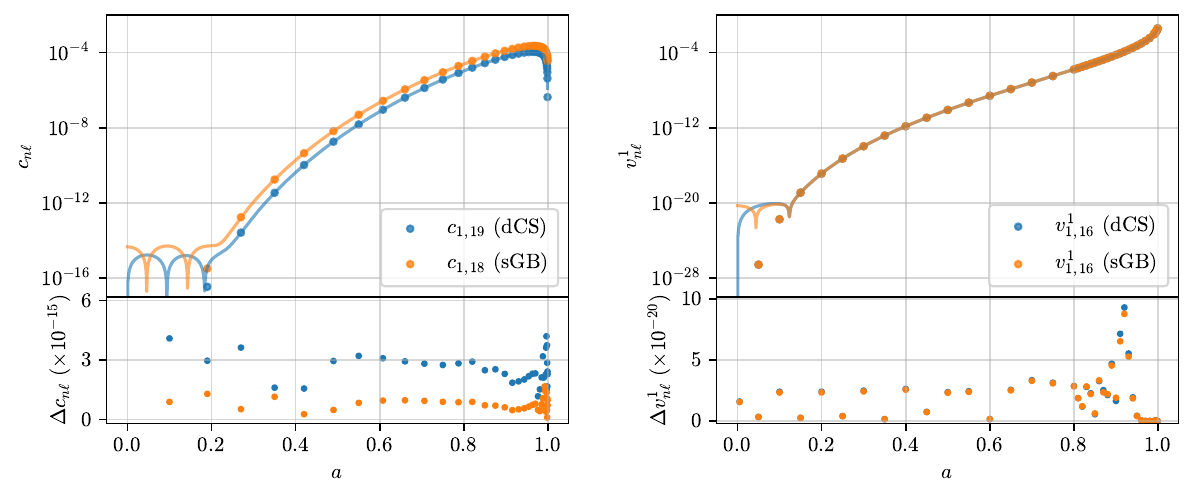}
    \caption{Top panels: Circle markers represent sGB (orange) and dCS (blue) spectral coefficients $c_{n\ell}$ (left) and $v_{n\ell}^i$ (right) of the spectral solution. 
    Solid lines connecting the markers represent the analytic, fitted spectral coefficients. 
    Bottom panel: Fitting residual $\Delta c_{n\ell}$ between the spectral solution and the analytic solution. 
    }
    \label{fig:SpectralCoefficients_Largest}
\end{figure*}

\begin{figure*}[t!]
    \centering
    \includegraphics[width=\textwidth]{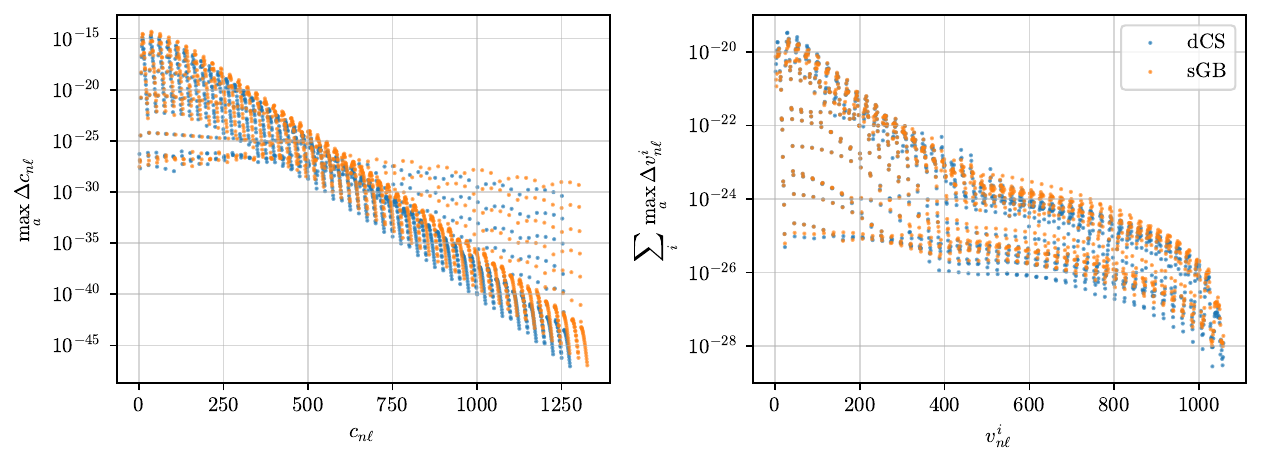}
    \caption{Maximum fitting residual of the sGB (orange) and dCS (blue) spectral coefficients $c_{n\ell}$ (left) and $v_{n\ell}^i$ (right). Observe that the maximum fitting residuals of all $c_{n\ell}$ ($v_{n\ell}^i$) are tiny, with order of magnitude $\lesssim 10^{-15}$ ($\lesssim 10^{-20}$), meaning the analytic solution is a faithful approximation of the spectral solution.}
    \label{fig:SpectralCoefficients_Residual}
\end{figure*}

We begin by recalling that the scalar field spectral solutions are fully characterized by $c_{n\ell}$ spectral coefficients, which can be used to reconstruct $\vartheta(r, \chi)$ through Eqs.~(\ref{eq:varthetaLegendreExpansion}) and (\ref{eq:vartheta_ell_expansion}). 
These spectral coefficients only depend on the BH spin, because the full radial and angular dependence is contained analytically in the Chebyshev and Legendre polynomial functions. The spectral coefficients, however, are obtained numerically, through a pseudospectral method described in Sec.~\ref{subsec:ScalarField}, at fixed values of BH spin.
By fitting these coefficients as a function of $a$, we can obtain a closed-form, analytic scalar field solution, with the full closed-form functional dependence on radius and polar angle encoded in the Chebyshev and Legendre polynomials.

To design an appropriate fitting ansatz, let us examine some of the spectral coefficients. 
In the upper left panel of Fig.~\ref{fig:SpectralCoefficients_Largest}, we show two spectral coefficients, $c_{1,19, \rm dCS}$ in blue and $c_{1,18, \rm sGB}$ in orange, as functions of spin. 
We show these two spectral coefficients because they capture the typical behaviors observed across all other spectral coefficients. 
For small $a$, the $c_{n\ell}$ coefficients are generally extremely small; as $a$ increases, these coefficients smoothly increase, and when $a \approx 0.9$, both coefficients reverse course sharply and begin to drop. 
To model this behavior, we find that a logarithmic ansatz that diverges at $a = 1$ gives the least fitting residual, $\Delta c_{n\ell} = |c_{n\ell, \rm fit} - c_{n\ell, \rm spectral}|$, with the least number of fitting coefficients required. 
The logarithmic ansatz also successfully captures the rapid change of behavior near $a \approx 0.9$. 
Other ansatze, such as polynomials and rational functions in $a$, often lead to numerical artifacts, e.g., spurious oscillations at small spins or unphysical poles within the domain of interest of $a$, and thus, we avoid using them here. 

More explicitly, we fit the spectral coefficients with the following functional forms:
\begin{equation}\label{eq:ScalarFieldFit}
\begin{split}
    c_{n\ell, \rm sGB} &= \sum_{m = 0}^{m_{\rm max}} \gamma_{n\ell,m} \left[\log(1 - a^2)\right]^m, \\
    c_{n\ell, \rm dCS} &= \sum_{m = 0}^{m_{\rm max}} \gamma_{n\ell,m} \, a \left[\log(1 - a^2)\right]^m, 
\end{split}
\end{equation}
where $\gamma_{n\ell,m}$ are fitting coefficients that are found using \emph{Mathematica}'s \texttt{NonlinearModelFit} function. 
Notice that we additionally include a factor of $a$ in the dCS fitting ansatz, which ensures that the $c_{n\ell}$ coefficients are odd functions of $a$.
For both scalar fields, we have taken approximately 250 spectral solutions (each of which contains $\sim 1300$ $c_{n\ell}$ coefficients at each value of spin), with increasing resolution near $a = 1$. This data is then fitted to the functions presented above to find $\gamma_{n\ell,m}$, setting $m_{\rm max} = 40$. 
Increasing $m_{\rm max}$ would result in a better fit, but the resultant fitting function becomes lengthy and not significantly better. 

\begin{figure}[t!]
    \centering
    \includegraphics[width=\columnwidth]{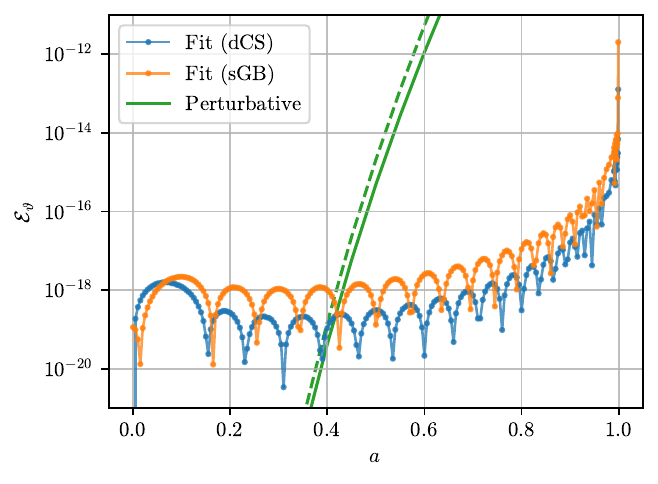}
    \caption{The absolute error of the sGB (orange) and dCS (blue) scalar field equation computed using the analytic solutions. The absolute error of the series-in-$a$ solutions up to ${\cal{O}}(a^{40})$, from Fig.~\ref{fig:ScalarFieldResidual50}, are also plotted in green for comparison. }
    \label{fig:ScalarFieldFit}
\end{figure}

In the left panels of Fig.~\ref{fig:SpectralCoefficients_Largest}, we also plot the fitted functions (and their residuals in the bottom panel) of the same two spectral coefficients as a function of $a$. 
In the top panel, observe that the fitting function overlaps with the spectral coefficients exceedingly well when $a > 0.2$. 
For smaller spins, we see that the markers no longer coincide with the analytic fit because the values of the spectral coefficient data become tiny (and in fact, it is comparable to the fitting residual of spectral coefficients at high spin). 
Therefore, \texttt{NonlinearModelFit} function cannot obtain a fit with great precision when the spin is that small. 
Despite this fitting deviation, $\Delta c_{n\ell}$ is generally very small over the entire range of $a$, with an order of magnitude of $10^{-15}$ for all spins fitted (even at $a < 0.2$), and thus, the analytic fit is still an excellent approximation to the spectral coefficients.  
Furthermore, $\Delta c_{n\ell}$ is approximately uniform for all spins, indicating that there are no specific regions of $a$ that fit better or worse than others. 

These two $c_{n\ell}$ coefficients exhibit the largest $\Delta c_{n\ell}$ relative to other coefficients, and hence, they represent the dominant sources of error.
To gauge the accuracy of all the fits, we show the maximum $\Delta c_{n\ell}$ over all values of $a$, denoted by $\max_a \Delta c_{n\ell}$, for all values of $n$ and $\ell$ in the left panel of Fig.~\ref{fig:SpectralCoefficients_Residual}. 
Observe that the fitting residual for all $c_{n\ell}$ is $\lesssim 10^{-15}$, almost reaching machine precision. 
This guarantees that the analytic solution we construct closely matches the spectral solution in the entire spacetime.

Now that we have verified that the fitting ansatz is capable of representing the spectral solution, we estimate the accuracy of the analytic scalar field solution by calculating the absolute error of the scalar field using the analytic fit via Eq.~(\ref{eq:ScalarFieldResidual}).
Figure~\ref{fig:ScalarFieldFit} shows the absolute error $\mathcal{E}_{\vartheta}$ of the analytic solution from $a = 0$ to $0.999$. 
For $a \leq 0.9$, $\mathcal{E}_{\vartheta}$ remains below machine precision. 
As $a$ increases, $\mathcal{E}_{\vartheta}$ reaches $10^{-12}$ at $a = 0.999$, consistent with the absolute error shown in Fig.~\ref{fig:ScalarFieldResidual50}. 
Compared to the series-in-$a$ solution, at $a \approx 0.4$, the analytic solution becomes more accurate than the series-in-$a$ solution.  
These results clearly show that our analytic solution is an excellent approximation to the true scalar field solution. 

%---------------------------------
\subsection{Metric Corrections}

\begin{figure}[t!]
    \centering
    \includegraphics[width=\columnwidth]{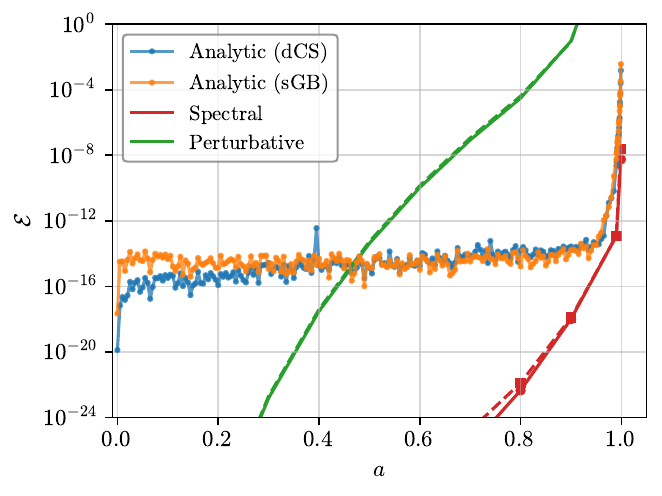}
    \caption{The absolute error of the modified Einstein field equations computed using the analytic scalar fields and metric corrections (orange and blue). For $a \leq 0.99$, the maximum absolute error is $\lesssim 10^{-8}$; for near extremal cases, the absolute error increases rapidly up to around $10^{-2}$ near $a = 0.999$. For direct comparison, the absolute error of the spectral solutions (red) and series-in-$a$ solutions (green) from Fig.~\ref{fig:EEResidual45} are displayed. }
    \label{fig:MetricCorrectionsFit}
\end{figure}

We now move on to the metric corrections and perform a similar analysis. In this case, the $v_{n\ell}^i$ coefficients fully determine the EFT-corrections to the metric. The radial and polar angle dependence of the metric is fully encoded in the Chebyshev and Legendre polynomials used in our spectral solution. The spectral coefficients, therefore, only depend on the BH spin $a$ and are calculated numerically for a fixed set of spin values. By constructing an analytic, closed-form fit for these spectral coefficients as a function of spin, we can therefore reconstruct the full exterior BH metric analytically.

Let us begin by studying the behavior of the spectral coefficients as a function of spin to gain some insight into their functional behavior. 
The upper, right panel of Fig.~\ref{fig:SpectralCoefficients_Largest} shows the spectral coefficients $v_{1,16}^1$ in dCS and sGB gravity as a function of spin.
These two coefficients capture the typical trends observed among all other spectral coefficients. 
Observe that as $a$ increases, the spectral coefficients increase rapidly toward $a = 1$. 
This behavior can again be captured by polynomials in $\log(1 - a^2)$, and thus, we choose the same fitting ansatz as for the scalar field, namely
\begin{equation}\label{eq:MetricCorrectionsFit}
    v_{n\ell}^{i} = \sum_{p = 0}^{p_{\rm max}} \nu_{n\ell,p}^{i} \, \left[\log(1 - a^2)\right]^p, 
\end{equation}
where the $\nu_{n\ell,p}^{i}$ are fitting coefficients. 
Since $H_i$ in both sGB and dCS theories are even functions of $a$, we omit the additional factor of $a$ we introduced when modeling the scalar field. 

Due to limited computational resources, computing hundreds of spectral BH solutions for various BH spin values to fit for the $\nu_{n\ell,p}^{i}$ is not feasible, so instead, we computed 45 different BH solutions with denser sampling near $a = 0.999$. 
Using this data, we then fit every spectral coefficient using \emph{Mathematica}'s \texttt{NonlinearModelFit} function, and we set $p_{\rm max} = 30$ to avoid overfitting. The top right panel of Fig.~\ref{fig:SpectralCoefficients_Largest} shows the resulting fitting function for the $v_{1,16}^1$ coefficients, we present excellent agreement to the data.

Let us now study the fitting residual of $v_{n\ell}^i$, $\Delta v_{n\ell}^i = |v_{n\ell, \rm fit}^i - v_{n\ell, \rm spectral}^i|$, as a function of $a$. 
In the bottom, right panel of Fig.~\ref{fig:SpectralCoefficients_Largest}, we show this residuals for the same two spectral coefficients. 
For small spins, although the fitted spectral coefficients do not overlap with the data points, the fitting residuals are tiny ($< 10^{-19}$). 
As such, the fit is still a good approximation of the spectral coefficients. 
To evaluate the overall fitting performance, we plot the sum of the maximum residual for all spectral coefficients, $\sum_i \max_a \Delta v_{n\ell}^i$ in Fig.~\ref{fig:SpectralCoefficients_Residual}. 
Observe that the maximum fitting residual is always below $10^{-19}$, ensuring that the analytic solution can represent the spectral solution. 

The absolute error of the field equations computed using the analytic solutions as a function of spin is shown in Fig.~\ref{fig:MetricCorrectionsFit}. 
For $a < 0.9$, the absolute error of the analytic metric corrections remains approximately constant at $\mathcal{E} \lesssim 10^{-14}$. 
Since the spectral coefficients $c_{n\ell}$ have a maximum fitting residual of $\sim 10^{-15}$, this translates to an absolute error of the field equation of around $10^{-15}$, limiting the accuracy the solution to roughly double precision.
We have tested that increasing the accuracy of the analytic scalar field, i.e.~increasing $m_{\rm max}$, leads to an overall smaller $\mathcal{E}$.
Nevertheless, the current analytic solution has been able to represent the spectral solution faithfully. 
Hence, we fix $m_{\rm max}$ at the current value to strike a balance between the length of the expression and computation resources needed and the accuracy of the solution.
At $a \approx 0.5$, the absolute error of the series-in-$a$ solution exceeds that of the analytic solution, indicating that the analytic solution is a better approximate solution to the modified field equations. 
Beyond $a = 0.9$, the absolute error increases rapidly.
This is to be expected since, when spin is large, the spectral solutions (red lines) have a relatively large absolute error. 
Despite this upsurge in error, the analytic solutions below $a = 0.99$ have $\mathcal{E} \lesssim 10^{-8}$. 
If one requires rapidly-rotating solutions with a higher accuracy, instead of using the analytic solution, one could always compute the spectral solution that guarantees an error below $10^{-8}$, as shown by the spectral solutions in red. However, as we show in the next section, observables calculated with our analytic metric are much more accurate than current instrumental and statistical errors associated with actual observations, so they should suffice. 

%----------------------------------------------------------------
\section{Properties of the Solution}\label{sec:PhysicalObservables}

In this section, we examine some properties of BHs in beyond-Einstein gravity using the spectral and analytic solutions found in the previous section. 
In view of studying the horizon behavior of BHs, we first discuss the leading-order EFT corrections of their surface gravity and their horizon angular velocity. 
We then compute the leading-order EFT corrections to the innermost stable circular orbit (ISCO) and the photon ring location, as this is relevant to geodesic motion in BH spacetimes outside Einstein's theory.  

Any physical quantity $A$ that we compute is to be expanded in the coupling constant of the EFT-corrected theory as
\begin{align}
    A = A^{(0)} + \zeta A^{(1)}\,,
\end{align}
where $A^{(0)}$ is the Kerr result for this observable, and $A^{(1)}$ is the leading-order EFT correction. Since $A^{(1)}$ is always multiplied by a factor of $\zeta \ll 1$, the EFT correction is always small compared to $A^{(0)}$ and can \textit{never} overwhelm the Kerr result.  
For any physical quantity $A$ that we compute (be it $A^{(0)}$ or $A^{(1)}$) using the analytic solution, $A_{\rm analytic}$, we can compare it to $A_{\rm series}$, the same physical quantity computed with the series-in-$a$ solution to ${\cal{O}}(a^{15})$. 
The difference can be quantified by the relative fractional error, defined as 
\begin{equation}\label{eq:PhysicalObservableRFE}
    \delta A = \left|1 - \frac{A_{\rm series}}{A_{\rm analytic}}\right|. 
\end{equation}
The fractional error of the series-in-$a$ solution allows us to determine the regime where the series-in-$a$ solution breaks down. 

\subsection{Surface Gravity}\label{subsec:SurfaceGravity}

For a stationary black hole with a timelike Killing vector $\xi_{(t)}^{\alpha}$, its surface gravity can be defined as \cite{Poisson:2009pwt}
\begin{align}
    \kappa \equiv - \frac{1}{2} (\nabla^{\beta} \xi_{(t)}^{\alpha}) (\nabla_{\beta} \xi^{(t)}_{\alpha})
\end{align}
evaluated at the event horizon $r=r_+$. The surface gravity can be interpreted as the force needed for an observer at spatial infinity to hold a test particle stationary at the horizon. For the Kerr metric, the surface gravity is simply $\kappa^{(0)} = (r_+ - M)/2Mr_+$. 
When we include the EFT-correction to the metric, as we have done in this paper, the leading-order correction to the surface gravity $\kappa^{(1)}$, defined through $\kappa = \kappa^{(0)} + \zeta \kappa^{(1)}$, is \cite{Cano_Ruiperez_2019}
\begin{equation}\label{eq:SurfaceGravity}
\begin{split}
    \kappa^{(1)} &= \frac{r_+ - M}{2Mr_+} \bigg[ H_2 - \frac{H_3}{2} - \frac{H_4}{2} + \frac{M^2 r_+^2}{(r_+ - M)\Sigma} \\
    & \dv{r}\left(-H_1 \Sigma + a^2(1-\chi^2)(2H_2 - H_4)\right) \\
    &+ 2(r_+ - M)(H_4 - 2H_2) \bigg] \bigg|_{r=r_+}. 
\end{split}
\end{equation}
By the zeroth-law of BH mechanics, which holds for any stationary BH in and outside general relativity~\cite{Racz:1992bp,Poisson:2009pwt}, we must have a constant (or uniform) surface gravity at the horizon. In particular, our BHs are also stationary, so $\kappa^{(1)}$ must also be constant over $\chi$, and any failure of this to hold represent error in our solution. 

Both the spectral and the analytic solutions we have constructed are extremely accurate (and can be systematically made more accurate by increasing the spectral order), but they are not formally exact. This is because the spectral $H_i$ are constructed by numerically inverting the matrix $\tilde{\mathbb{D}}$ to obtain the spectral expansion coefficients at a given spin, and this numerical inversion brings in numerical error (which is then also inherited by the analytic solution). We have studied this before, when evaluating the $L^2$ norms, but we can also gauge this error through the $\chi$ dependence of the surface gravity. To do so, let us take the value of $\kappa^{(1)}$ at the equator as a measure of the ``true'' value, and compute the numerical ``error'' of $\kappa^{(1)}$ along $\chi$, defined by 
\begin{equation}\label{eq:SurfaceGravityChiError}
    \left|\!\left|\frac{d\kappa^{(1)}}{d\chi}\right|\!\right|_2 = \left[ \int_{-1}^{+1} \left(\dv{\chi}\kappa^{(1)}\right)^2 \,d\chi \right]^{1/2}.
\end{equation}
If $\kappa^{(1)}$ is independent of $\chi$, then $|\!|d\kappa^{(1)}/d\chi|\!|_2$ vanishes exactly. 
Hence, $|\!|d\kappa^{(1)}/d\chi|\!|_2$ quantitatively measures the $\chi$ dependence of $\kappa^{(1)}$, and thus, the error in our solution. 

\begin{figure}[t!]
    \centering
    \includegraphics[width=\columnwidth]{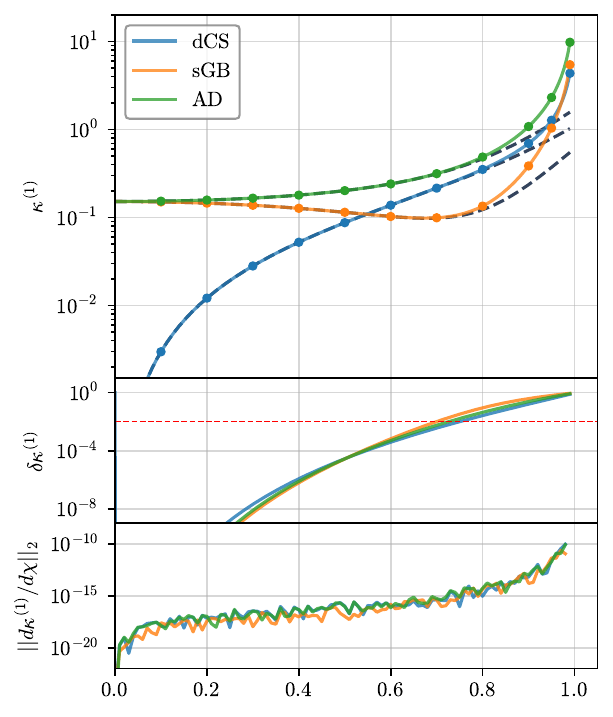}
    \caption{The correction to the surface gravity and its error estimates as a function of spin up to $a = 0.99$ in dCS, sGB and AD gravity on the equator. 
    In the top panel, markers represent $\kappa^{(1)}$ computed using Eq.~(\ref{eq:SurfaceGravity}) and the spectral solutions, while the solid lines connecting are $\kappa^{(1)}$ computed with analytic solutions. The gray dashed lines show $\kappa^{(1)}$ computed with the series-in-$a$ solution up to ${\cal{O}}(a^{15})$. 
    The middle panel shows the relative fractional error $\delta \kappa^{(1)}$ between the analytic solution and series-in-$a$ solution (gray dashed line), with red horizontal dashed line indicating $\delta \kappa^{(1)} = 1\%$.  Observe that the series-in-$a$ solution begins to introduce error (above the 1\% level,  relative to the spectral and the analytic solution) once $a \gtrsim 0.7$. 
    The bottom panel displays the error measure $|\!|d\kappa^{(1)}/d\chi|\!|_2$, as defined in Eq.~(\ref{eq:SurfaceGravityChiError}), against $a$.
    Observe that the numerical and the analytic solutions both have similar errors, with a maximum of $10^{-10}$ at a spin of $0.99$. 
   }
    \label{fig:SurfaceGravity}
\end{figure}

The top panel of Fig.~\ref{fig:SurfaceGravity} shows $\kappa^{(1)}$ as a function of $a$ computed with the spectral, the analytical and the series-in-$a$ solution to ${\cal{O}}(a^{15})$ for comparison.
Observe that the correction to the surface gravity is always positive, indicating that a larger force is needed to keep a test particle stationary at the event horizon. 
Observe also that the dCS correction goes to zero when $a = 0$, which is consistent with the fact that the Schwarzschild metric is a solution in dCS gravity.

The $\kappa^{(1)}$ computed with the spectral solution agrees with that computed with the analytic solution extremely well, but differs with that computed with the series-in-$a$ solution at large spins. 
More concretely, the EFT corrections to the surface gravity computed with all three metrics agree up to $a \approx 0.8$.
When $a > 0.8$, however, the $\kappa^{(1)}$ computed with the spectral and analytic solutions agree with each other but deviate from that computed with the series-in-$a$ solution, showing that the latter fails to accurately capture the properties of BHs with large spins. We can study this difference more clearly through the middle panel of Fig.~\ref{fig:SurfaceGravity}, which shows the relative fractional difference $\delta \kappa^{(1)}$ with respect to $\kappa^{(1)}$ computed with the series-in-$a$ solution.
Observe that $\delta \kappa^{(1)}$ increases with $a$, reaching $1\%$ near $a = 0.7$, as shown by the red dashed line. 
This agrees with the estimates of~\cite{Cano_Ruiperez_2019}, whose series-in-$a$ solution is only valid up to $a = 0.7$ if one demands a 1\% accuracy. 

The bottom panel of Fig.~\ref{fig:SurfaceGravity} shows the error measure $|\!|d\kappa^{(1)}/d\chi|\!|_2$ for the spectral and the analytic solutions.
Observe that $|\!|d\kappa^{(1)}/d\chi|\!|_2$ is below machine precision for solutions with small to moderate spins, indicating that $\kappa^{(1)} $ is indeed numerically uniform across $\chi$, and that the metric satisfies the zeroth-law of BH mechanics. 
As one increases $a$, $|\!|d\kappa^{(1)}/d\chi|\!|_2$ rises to $\sim 10^{-10}$ at $a = 0.99$, indicating the level of precision of the solution and the observable.  In particular, at spins of $0.7$ the error measure $|\!|d\kappa^{(1)}/d\chi|\!|_2$ is approximately double precision.  

\subsection{Horizon Angular Velocity}

\begin{figure}[t!]
    \centering
    \includegraphics[width=\columnwidth]{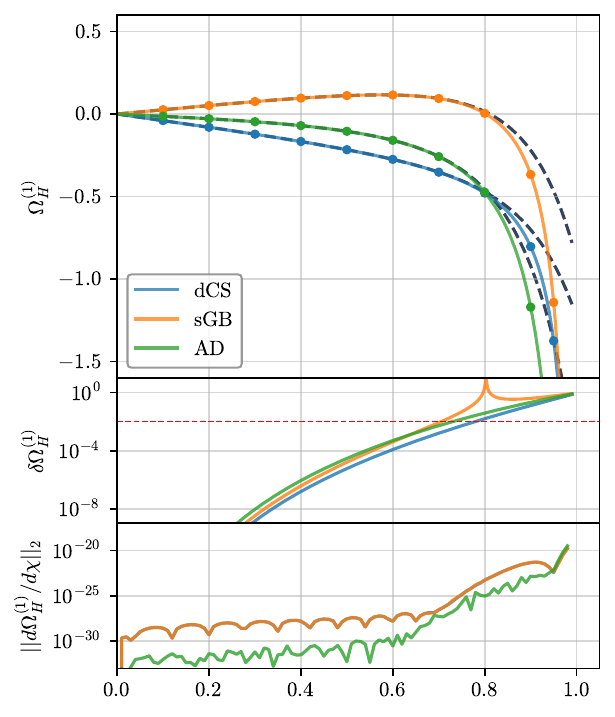}
    \caption{Same as Fig.~\ref{fig:SurfaceGravity}, except that the observable is the horizon angular velocity correction $\Omega_H^{(1)}$, defined by Eq.~(\ref{eq:Omega_H}), on the equator.}
    \label{fig:HorizonAngularVelocity}
\end{figure}

Next, we compute the horizon angular velocity $\Omega_H$ in sGB, dCS and AD gravity. For a stationary and axisymmetric black hole, the horizon angular velocity is defined through a ratio of components of the metric in Boyer-Lindquist coordinates, $\Omega_H \equiv |g_{t \phi}|/g_{\phi\phi}$, evaluated at the event horizon $r_+$. Physically, this quantity represents the angular velocity that stationary observers at the horizon would have~\cite{Poisson:2009pwt}. For a Kerr BH, $\Omega_H^{(0)} = a/(2Mr_+)$, and thus, the horizon angular velocity correction $\Omega_H^{(1)}$, defined by $\Omega_H = \Omega_H^{(0)} + \zeta \Omega_H^{(1)}$, can be found to be \cite{Cano_Ruiperez_2019}
\begin{equation}\label{eq:Omega_H}
    \Omega_H^{(1)} = \frac{a}{2Mr_+}\Big(H_2 - H_4\Big)\Big|_{r=r_+}.
\end{equation}

Just as in the case of the surface gravity, one expects the horizon angular velocity to be independent of polar angle, and thus, independent of $\chi$ \cite{Poisson:2009pwt}. The dependence of this function on $\chi$ is therefore a measure of the error in our spectral solution, just as in the case of $\kappa^{(1)}$. Let us then take $\Omega_H^{(1)}$ at the equator as the ``true'' value, and measure the error in our solution along $\chi$ by 
\begin{equation}\label{eq:HorizonAngularVelocityChiError}
    \left|\!\left|\frac{d\Omega_H^{(1)}}{d\chi}\right|\!\right|_2 = \left[ \int_{-1}^{+1} \left(\dv{\chi} \Omega_H^{(1)}\right)^2 \,d\chi \right]^{1/2}. 
\end{equation}
For an exact solution, we expect $|\!|d\Omega_H^{(1)}/d\chi|\!|_2$ to vanish exactly.  

Figure~\ref{fig:HorizonAngularVelocity} shows the horizon angular velocity of the EFT-corrected BH solutions computed with the spectral solutions (markers), analytic solutions (solid lines), series-in-$a$ solutions (gray dashed lines) in the top panel. Observe that the dCS and AD corrections tend to slow down the angular velocity of stationary observers on the event horizon. All other features are similar to those of the surface gravity: the $\Omega_H^{(1)}$ computed with the spectral and analytic solutions agree with each other, but disagree with the series-in-$a$ solution once $a \gtrsim 0.7$. This disagreement can be seen better in the middle panel, where we show the relative fractional difference in $\Omega_H^{(1)}$ computed with the spectral or the analytic resolutions and the series-in-$a$ solution. 

The bottom panel of Fig.~\ref{fig:HorizonAngularVelocity} shows the error measure $|\!|d\Omega_H^{(1)}/d\chi|\!|_2$ as a function of spin. 
For all spins, the error measure $|\!|d\Omega_H^{(1)}/d\chi|\!|_2$ is below machine precision, ensuring that $\Omega_H^{(1)}$ is numerically uniform across $\chi$ on the horizon. 
Observe also that $|\!|d\Omega_H^{(1)}/d\chi|\!|_2$ is much smaller than $|\!|d\kappa^{(1)}/d\chi|\!|_2$, especially near the extremal limit. 
Since the expression for $\kappa^{(1)}$ in Eq.~\eqref{eq:SurfaceGravity} contains a factor of $\Sigma^{-1}$, it is more difficult for $\kappa^{(1)}$ to be fully $\chi$ independent; the same is not true for the expression for $\Omega_H^{(1)}$ in Eq.~\eqref{eq:Omega_H}, which is manifestly $\Sigma$ independent. In fact, we could demand that $\frac{\partial H_2}{\partial \chi}(r_+, \chi) = \frac{\partial H_4}{\partial \chi}(r_+, \chi)$ as an additional constraint when solving for the spectral expansion coefficients $v_{n \ell}^i$; we have checked that doing so would lead to $|\!|d\Omega_H^{(1)}/d\chi|\!|_2 = 0$ exactly, but yet, the error measure ${\cal{E}}$ in Eq.~\eqref{eq:EEResidual} would remain roughly the same order of magnitude.

\subsection{Innermost stable circular orbit}

\begin{figure*}[t!]
    \centering
    \includegraphics[width=\textwidth]{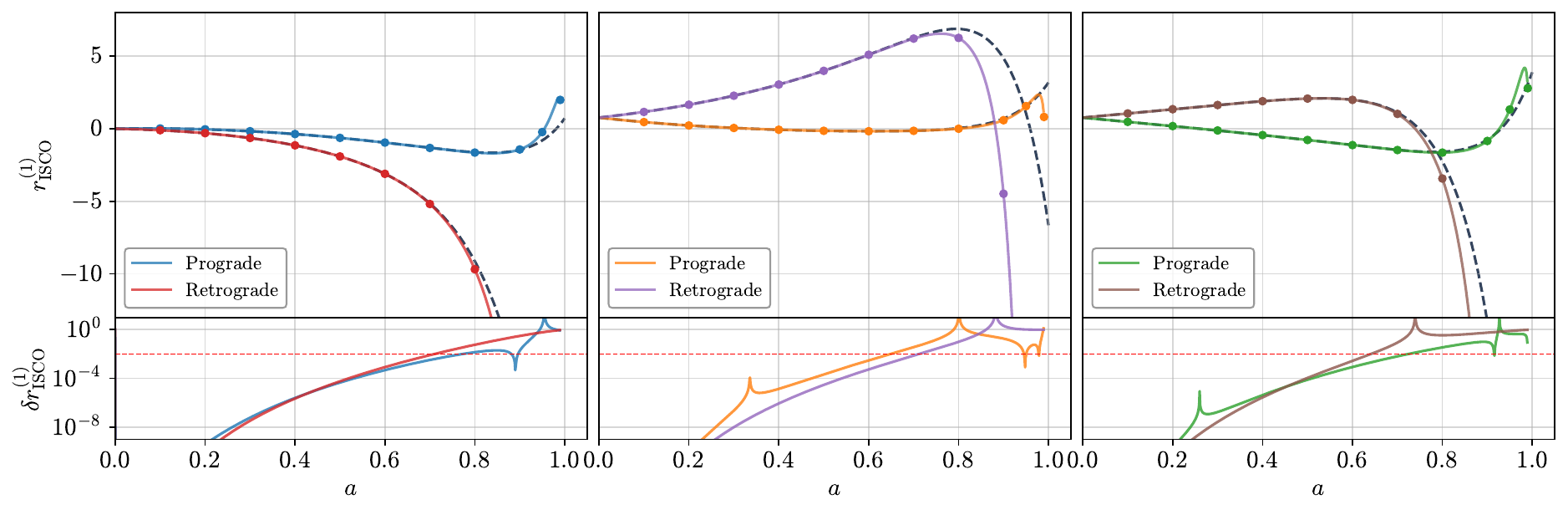}
    \caption{Top panel: ISCO radii corrections in dCS (left), sGB (middle) and AD (right) gravity as a function of spin up to $a = 0.99$. 
    The markers show $r_{\rm ISCO}^{(1)}$ computed with spectral solutions, while the solid lines represent $r_{\rm ISCO}^{(1)}$ computed with the analytic solution. 
    The gray dashed lines show $r_{\rm ISCO}^{(1)}$ computed with the series-in-$a$ solution up to 15-th order in $a$. 
    Bottom panel: The relative fractional error $\delta r_{\rm ISCO}^{(1)}$ between the analytic solution and the series-in-$a$ solution, with red dashed line indicating $\delta r_{\rm ISCO}^{(1)} = 1\%$. }
    \label{fig:ISCO}
\end{figure*}

For any stationary and axisymmetric spacetime, the Killing vectors $\xi_{\mu}^{(t)}$ and $\xi_{\mu}^{(\phi)}$ generate two constants of geodesic motion for test particles, $E$ and $L$, respectively \cite{Yagi:2012ya, Cano:2023qqm}. 
Using their definitions, we can obtain 
\begin{equation}
    \dot{t} = \frac{E g_{\phi\phi} + L g_{t\phi}}{g_{t\phi}^2 - g_{tt}g_{\phi\phi}}, \qquad 
    \dot{\phi} = -\frac{E g_{t\phi} + L g_{tt}}{g_{t\phi}^2 - g_{tt}g_{\phi\phi}}. 
\end{equation}
For timelike geodesics, the particle's 4-velocity $u^{\mu}$ is normalized, i.e., $u^{\mu}u_{\mu} = -1$, and thus,  
\begin{equation}
    g_{rr}\dot{r}^2 + g_{\chi\chi}\dot{\chi}^2 = V_{\rm eff}(r, \chi), 
\end{equation}
where 
\begin{equation}
    V_{\rm eff}(r, \chi) = \frac{E^2 g_{\phi\phi} + 2EL g_{t\phi} + L^2 g_{tt}}{g_{t\phi}^2 - g_{tt}g_{\phi\phi}} - 1
\end{equation}
is the effective potential. 
Since reflection symmetry is enforced in our solution, orbits contained in the equatorial plane remain in the equatorial plane  \cite{Cano_Ruiperez_2019}. 

Let us now focus on equatorial and circular orbits, which implies $V_{\rm eff}(r) = 0$ and $V'_{\rm eff}(r) = 0$. 
This allows us to find the corrections to $E(r)$ and $L(r)$, $E^{(1)}(r)$ and $L^{(1)}(r)$, defined as 
\begin{equation}
\begin{split}
    E(r) = E^{(0)}(r) + \zeta E^{(1)}(r) \\
    L(r) = L^{(0)}(r) + \zeta L^{(1)}(r),
\end{split}
\end{equation}
where 
\begin{equation}
\begin{split}
    E^{(0)}(r) &= \frac{r^{3/2} - 2Mr^{1/2} + aM^{1/2}}{r^{3/4}(r^{3/2} - 3Mr^{1/2} + 2aM^{1/2})^{1/2}}, \\
    L^{(0)}(r) &= \frac{M^{1/2}(r^{2} - 2aM^{1/2}r^{1/2} + a^{2})}{r^{3/4}(r^{3/2} - 3Mr^{1/2} + 2aM^{1/2})^{1/2}}, \\
\end{split}
\end{equation}
are the specific energy and angular momentum of a massive test particle in the Kerr spacetime \cite{Bardeen:1972fi}. 
The energy and angular momentum corrections can be solved for in terms of $H_i$ and zeroth order quantities, but the expressions are long and not insightful, and thus, we do not present them here. 

To compute the location of the ISCO $r_{\rm ISCO}$, we further impose the marginally stable condition $V''_{\rm eff}(r_{\rm ISCO}) = 0$. Doing so, we can solve for the EFT correction to the location of the ISCO $[r_{\rm ISCO}^{\pm}]^{(1)}$, defined by 
\begin{equation}
    r_{\rm ISCO}^{\pm} = [r_{\rm ISCO}^{\pm}]^{(0)} + \zeta [r_{\rm ISCO}^{\pm}]^{(1)},
\end{equation}
where 
\begin{equation}
    [r_{\rm ISCO}^{\pm}]^{(0)} = M\left[3 + Z_2 \mp \sqrt{(3 - Z_1)(3 + Z_1 + 2Z_2)}\right], 
\end{equation}
$Z_1 = 1 + (1 - a^2)^{1/3} [(1 + a)^{1/3} + (1-a)^{1/3}]$ and $Z_2 = (3a^2 + Z_1^2)^{1/2}$ \cite{Bardeen:1972fi}. The $\pm$ sign represents prograde and retrograde orbits respectively.

Figure~\ref{fig:ISCO} shows the EFT corrections to the location of the ISCO for both prograde and retrograde orbits in the equatorial plane\footnote{These corrections are not the same as the ``perimeter radius'' shown in \cite{Cano_Ruiperez_2019, Cano:2023qqm}.}.
Perhaps interestingly, as we increase $a$, the ISCO corrections to prograde orbits in sGB gravity reverse course and become negative rapidly. 
As before, $r_{\rm ISCO}^{(1)}$ computed with the spectral and analytic solutions agree with each other, and they disagree with that computed with the series-in-$a$ solution for spins $a \gtrsim 0.7$. This behavior can be seen clearly from the bottom panels of Fig.~\ref{fig:ISCO}, where we compare the spectral solution to the series-in-$a$ solution.

\subsection{Photon Ring}

\begin{figure*}[t!]
    \centering
    \includegraphics[width=\textwidth]{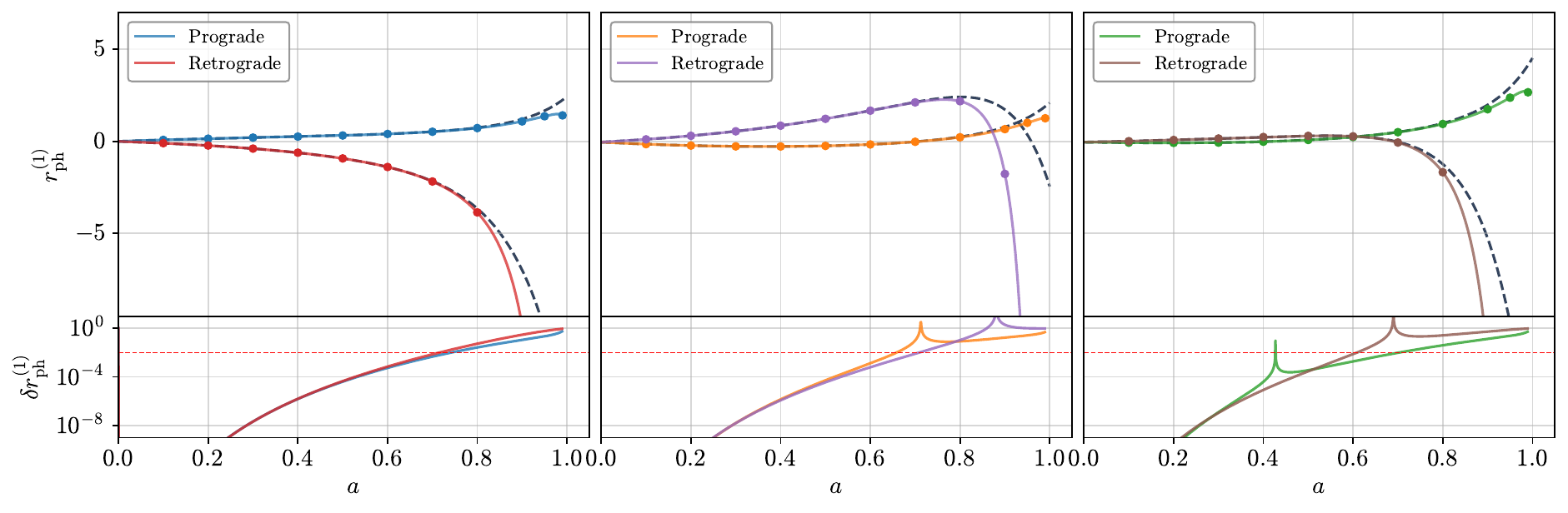}
    \caption{Same as Fig.~\ref{fig:ISCO}, except that the observable is the photon ring radii correction $r_{\rm ph}^{(1)}$. }
    \label{fig:PhotonRing}
\end{figure*}

The final observable that we study here is the photon ring radius. For simplicity, we restrict attention to equatorial, circular orbits. Following \cite{Cano_Ruiperez_2019}, we solve the geodesic equation and normalization constraint simultaneously to calculate the photon ring radius corrections, $[r_{\rm ph}^{\pm}]^{(1)}$, defined by 
\begin{equation}
    r_{\rm ph}^{\pm} = [r_{\rm ph}^{\pm}]^{(0)} + \zeta [r_{\rm ph}^{\pm}]^{(1)}, 
\end{equation}
where \cite{Bardeen:1972fi}
\begin{equation}
    [r_{\rm ph}^{\pm}]^{(0)} = 2M\left[1 + \cos(\frac{2}{3}\arccos(\mp a/M))\right],
\end{equation}
is the photon ring radius in GR. 
The $\pm$ sign denotes prograde and retrograde orbits respectively.

Figure~\ref{fig:PhotonRing} shows the EFT corrections to the location of the photon ring as a function of spin\footnote{As in the case of the ISCO, we focus here on the radial coordinate corrections, instead of the perimeter radius corrections~\cite{Cano_Ruiperez_2019}.}.
Observe that for prograde orbits, the EFT-corrections increase the size of the photon ring. As in the ISCO case, the sGB correction reverses direction as the spin increases, shrinking the photon ring.
As in all other cases, the EFT corrections to the location of the photon ring agree when computed with the spectral and the analytic solutions, but they disagree with those computed with the series-in-$a$ solution at spins $a \gtrsim 0.7$. This can again be seen more clearly in the bottom panels of Fig.~\ref{fig:PhotonRing}, where we show the relative fractional difference between $\delta r_{\rm ph}^{(1)}$ computed with the spectral or analytic solution and that computed with the series-in-$a$ solution. 

%----------------------------------------------------------------
\section{Conclusion}\label{sec:Conclusion}

We have developed an efficient and highly accurate framework to compute rotating black-hole (stationary, axisymmetric, vacuum) solutions in a large class of beyond-Einstein gravity theories through spectral expansions. These solutions are represented in terms of analytic and closed form functions of the spacetime coordinates, and they are valid to leading order in the small coupling expansion (consistent with the fact that these beyond-Einstein theories are EFTs) and to all orders in spin. 
To exemplify this framework, we applied it to scalar Gauss-Bonnet, dynamical Chern-Simons, and axi-dilaton gravity, thus obtaining the most accurate and the first closed-form and analytic spinning black hole solutions in these theories (all of which are provided as supplementary material in a \emph{Mathematica} notebook). We have explored the accuracy and the physical properties of these solutions for black holes with spins $a \in [0,0.99]$, focusing particularly on their surface gravity, horizon angular velocity, ISCO and photon ring locations. 

The solutions presented here (especially the analytic ones) open the road to a plethora of future investigations because they are valid at astrophysically-relevant spins. 
For example, the analyses of gravitational-wave observations from the mergers of stellar-mass black holes and electromagnetic observations of supermassive black holes require the use of accurate black hole metrics with which to calculate observables that can be compared to data~\cite{Ayzenberg:2023hfw, Tamburini:2019vrf}. 
The metrics we have calculated can be used to compute observables in EFT extensions of general relativity, thus enabling robust tests of Einstein's theory with gravitational-wave and electromagnetic observations. 
As a demonstration, we have computed some physical observables of a set of EFT-corrected black holes that are potentially applicable in these tests, such as the location of the ISCO and the photon ring. 

Another direct application of the spectral solution (and its analytic fit) is to compute black hole quasinormal modes in quadratic gravity, following, for example, the METRICS approach developed in~\cite{Chung:2023zdq, Chung:2023wkd, Chung:2024vaf, Chung:2024ira, Chung:2025gyg}. 
The accurate black hole backgrounds computed here should allow us to compute the quasinormal spectra of rapidly rotating black holes, especially those with spins $a > 0.85$ ($a > 0.75$) for scalar Gauss-Bonnet (dynamical Chern-Simons) gravity. 
Doing so will provide a complete quasinormal spectra that could then be used in gravitational-wave ringdown tests with a variety of publicly-available gravitational-wave data \cite{LIGOScientific:2025slb, LIGOScientific:2021djp, LIGOScientific:2020ibl}. 

The spectral and analytic solutions found here can also serve as a background spacetime to evolve extreme-mass-ratio inspirals.
Since these events are expected to be observed with future gravitational-wave missions, such as LISA \cite{Gair:2017ynp, Chamberlain:2017fjl, Babak:2017tow}, the metrics obtained here could enable new tests of general relativity around rapidly spinning black holes.
In this context, the spectral and analytic solutions we obtained will be valuable for probing quadratic gravity theories.
Furthermore, these solutions can also provide rapidly-rotating, black-hole initial data for numerical relativity simulations in quadratic gravity theories  \cite{Richards:2025ows, Okounkova_Scheel_Teukolsky_2019, Witek:2020uzz}. 

Compared to the series-in-$a$ solution constructed using slow-rotation expansion, our spectral and analytic solutions display a significant improvement in accuracy and efficiency. 
When spin is small, the spectral solution is in good agreement with the series-in-$a$ solution. 
However, for spins $a \gtrsim 0.7$, our solutions deviate from the series-in-$a$ ones, revealing the failure of the slow-rotation expansion. 
Since spectral methods do not require any spin expansions, our method remains accurate and efficient for very high spins. 

Future work could also focus on further improving the framework presented here. One such path is to improve the framework for extremal and near extremal black holes. 
The asymptotic behavior of the metric in the extremal limit, such as divergences near the horizon \cite{Chen:2018jed, Kleihaus:2015aje}, might need to be peeled off to improve the framework. Analytic knowledge of the behavior of the scalar field (and the trace of the metric perturbations) in the extremal limit for some quadratic gravity theories~\cite{McNees:2015srl}, for example, could be folded into the scalar field (and metric) ansatz. The behavior of certain observables at or near the horizon of extremal black holes may also signal the breakdown of EFTs, as studied for other EFT-corrected black holes in~\cite{Horowitz:2023xyl, Horowitz:2024dch}. 

Lastly, the framework presented here and in \cite{Lam:2025elw} is general enough that it can be applied to many other beyond-Einstein theories, such as cubic or quartic gravity \cite{Bueno:2016xff, Hennigar:2016gkm, Ahmed:2017jod}, theories with massive fields~\cite{Doneva:2019vuh,Li:2025ffh}, or theories with other coupling functions~\cite{Silva:2017uqg,Mignemi:1992nt, Moura:2006pz}. 
If the beyond-Einstein theory smoothly deforms black holes from the Kerr metric\footnote{This condition implies that black hole solutions that are on a branch that does not smoothly connect to Kerr, such as scalarized black holes in dCS and Einstein-dilaton-Gauss-Bonnet gravity \cite{Doneva:2021dcc, Doneva:2022ewd}, cannot be captured by our framework.}, one should be able to apply our framework. In this way, the work presented here drastically improves our ability to explore the landscape of beyond-Einstein black holes and carry out tests of general relativity with observational data. 

%----------------------------------------------------------------
\section*{Acknowledgment}

The authors would like to thank Stef Husken for identifying typographical errors in our manuscript.
The authors acknowledge the support from the Simons Foundation through Award No. 896696, the Simons Foundation International under grant SFI-MPS-BH-00012593-01, the NSF through Grant No. PHY-2207650 and PHY 25-12423, and NASA through Grant No. 80NSSC22K0806. 
A. K. W. C also acknowledge the Herchel Smith Fellowship at the University of Cambridge for partial support of this work. 
The calculations and results reported in this paper were produced using the computational resources of the Illinois Campus Cluster, a computing resource that is operated by the Illinois Campus Cluster Program (ICCP) in conjunction with National Center for Supercomputing Applications (NCSA), and is supported by funds from the University of Illinois at Urbana-Champaign.
The author would like to specially thank the investors of the IlliniComputes initiatives and GravityTheory computational nodes for permitting the authors to execute runs related to this work using the relevant computational resources. 

\bibliography{ref}

\end{document}